# A Printed Microscopic Universal Gradient Interface for Super Stretchable Strain-Insensitive Bioelectronics

*Kaidong Song, Jingyuan Zhou, Chen Wei, Ashok Ponnuchamy, Md Omarsany Bappy, Yuxuan Liao, Qiang Jiang, Yipu Du, Connor J. Evans, Brian C. Wyatt, Thomas O' Sullivan, Ryan K. Roeder, Babak Anasori, Anthony J. Hoffman, Lihua Jin, Xiangfeng Duan\* and Yanliang Zhang\**

K. Song, M. O. Bappy, Y. Liao, Q. Jiang, Y. Du, C. Evans, R. Roeder, Y. Zhang
Department of Aerospace and Mechanical Engineering
University of Notre Dame
Notre Dame, IN 46556, USA
E-mail: yzhang45@nd.edu

J. Zhou, X. Duan
Chemistry and Biochemistry Department
University of California Los Angeles
Los Angeles, CA 90095, USA
E-mail: xduan@chem.ucla.edu

C. Wei, L. Jin
Department of Mechanical and Aerospace Engineering
University of California Los Angeles
Los Angeles, CA 90095, USA

A. Ponnuchamy, Thomas O' Sullivan A. Hoffman
Department of Electrical Engineering
University of Notre Dame
Notre Dame, IN 46556, USA

B. Wyatt, B. Anasori
School of Materials Engineering
Purdue University
West Lafayette, IN 47907, USA

**Abstract**

Stretchable electronics capable of conforming to nonplanar and dynamic human body surfaces are central for creating implantable and on-skin devices for high-fidelity monitoring of diverse physiological signals. While various strategies have been developed to produce stretchable devices, the signals collected from such devices are often highly sensitive to local strain, resulting in inevitable convolution with surface strain-induced motion artifacts that are difficult to distinguish from intrinsic physiological signals. Here we report all-printed super stretchable strain-insensitive bioelectronics using a unique universal gradient interface (UGI) to bridge the gap between soft biomaterials and stiff electronic materials. Leveraging a versatile aerosol-based multi-materials printing technique that allows precise spatial control over the local stiffnesses with submicron resolution, the UGI enables strain-insensitive electronic devices with negligible resistivity changes under a 180% stretch ratio. We demonstrate various stretchable devices directly printed on the UGI for on-skin health monitoring with high signal quality and near perfect immunity to motion artifacts, including semiconductor-based photodetectors for sensing blood oxygen saturation levels and metal-based temperature sensors. The concept in this work will significantly simplify the fabrication and accelerate the development of a broad range of wearable and implantable bioelectronics for real-time health monitoring and personalized therapeutics.

# 1. Introduction

Stretchable bioelectronics hold the key for creating implantable and on-skin devices that can conform to nonplanar and dynamic human body surfaces[1,2], which are indispensable for high-fidelity continuous monitoring of physiological signals and personalized healthcare [3,4]. A major challenge in these emerging technologies is the significant mismatch between the properties of soft biological systems and stiff electronic devices. To realize stable integration with biological tissues with seamless conformal interfaces, it is crucial to ensure that the physiological deformation of soft tissues is not impeded by the bioelectronics and that the deformation of the tissues does not interfere or impair the functionality of the electronic devices.

To enhance the stretchability of electronic devices, considerable efforts have been devoted to the development of intrinsically stretchable materials that preserve electrical connectivity under extensive deformation[5–12]. However, the electronic properties of these stretchable materials are generally inferior to the conventional non-stretchable materials, and their properties often drift during repeated stretching cycles[13,14]. Alternatively, geometrical design strategies (e.g., serpentines and wrinkles) have been employed to enhance the stretchability of thin metal films by converting strain into structural bending, achieving high conductivity and stability within a specific strain limit[6,15,16]. However, these designs demand specialized fabrication techniques for patterning on elastomeric substrates and extra space for the necessary bends and folds, potentially leading to larger devices and lower packing density of components, reducing areal efficiency[17]. Additionally, the microscopic wrinkles often prevent the conformal interface that is critical for signal transduction.

Another strategy for stretchable electronics is to generate a nonuniform strain distribution by using substrates comprising strain-free and strain-absorbing regions[18,19]. Recent developments in strain-engineered elastomeric substrates using island-bridge structures have been employed for stretchable electronics[4,20]. Such designs include embedding stiff platforms within soft substrates[9,21–23], introducing soft interlayer[24], utilizing core/shell packages with ultralow modulus cores[25], and incorporating liquid-filled cavities in the soft substrate[26] to protect functional components from strain. Nonetheless, these systems all involve stiff-soft interfaces with a sharp change of mechanical properties, which are prone to mechanical failures due to high stress concentrations[22,23,27]. To this end, it is essential to design interfaces that can bridge the gap between stiff components and soft surfaces, ensuring they can withstand strains without compromising performances.

Nature provides an elegant solution to bridge the gap between soft and stiff materials via a functionally graded interface of gradient stiffness, allowing stiff components to adhere intimately with soft materials while minimizing the interfacial stresses and avoiding delamination. This strategy has been adopted to produce functionally graded interfaces via various fabrication techniques such as adjusting polymer chain cross-linking degree by tuning the photocuring[28], solvent-welding of multiple layers with varying reinforcement levels[27], multi-material extrusion printing[29], and adjusting local phase transitions in hydrogels[30]. However, the majority of these mechanically graded materials only have one-dimensional property gradient and relatively course stiffness gradient resolution (i.e. millimeter scale). Major challenges still exist in fabricating these interfaces with smooth gradient stiffness distributions with high spatial resolution and realizing precise control over the local compositions and properties across all the three dimensions.

Herein, we report all-printed super stretchable strain-insensitive bioelectronics using a unique universal gradient interface (UGI) to bridge the gap between soft tissues and stiff electronic materials. As shown in **Figure 1**, the UGI with submicron stiffness gradient is printed using an aerosol-based multi-materials printing (AMMP) by in-situ modulating the mixing ratio of multiple inks of vastly different properties in a single nozzle, realizing precise control of the local properties along all the three dimensions. A variety of semiconducting and metallic 2D/1D/0D nanomaterials can be printed on the UGI to form electronics with remarkable strain-insensitivity and negligible resistivity changes under high stretch ratio up to 180%. The all-printed devices deliver high signal quality and near perfect immunity to motion artifacts for on-skin health monitoring of blood oxygen saturation levels, temperature and pulses. Our facile approach greatly simplifies the fabrication of super-stretchable bioelectronics, enabling harmonious integration of a broad range of functional devices with biological systems for emerging applications such as wearable/implantable devices, soft/hybrid robotics, and human/machine interfaces, etc.

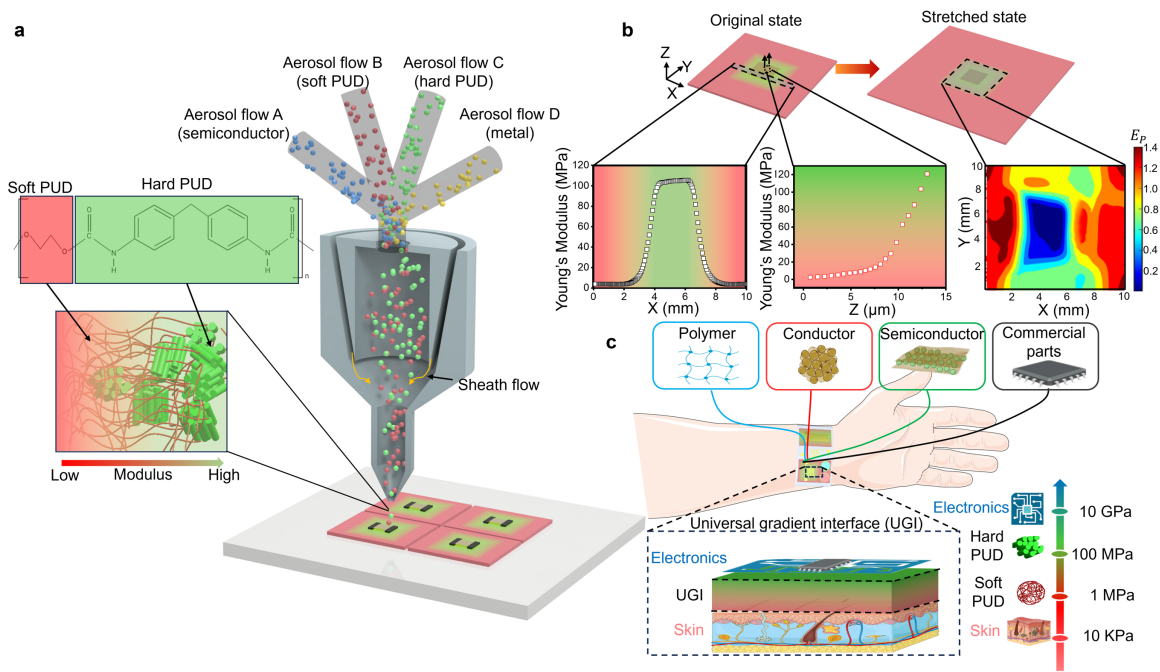

**Figure 1. Stretchable devices enabled by universal gradient interfaces (UGIs) printed using aerosol-based multi-materials printing (AMMP). a)** Scheme for one-step AMMP to print both UGI and functional devices. **b)** Schematic illustration of the original and stretched states of a 3D gradient UGI. Young's modulus distributions of the UGI and the maximum principal Lagrange strain map of the 3D UGI under 100% bi-axial stretch ratio. **c)** Schematic illustration of the wide range of materials that can be integrated with the UGI.

## 2. Universal Gradient Interfaces (UGIs) using AMMP

AMMP allows to directly print the UGI with gradient modulus with submicron spatial resolution. The AMMP approach starts by atomizing two (or multiple) inks into aerosols that consist of microscale droplets. These aerosols are then mixed in-situ in a single nozzle and focused using a co-flowing sheath gas prior to deposition (**Figure 1a**). The AMMP enables precise control over local material composition by dynamically modulating the mixing ratio of different aerosolized inks. Taking advantage of our multi-material printing process, the UGI and functional devices can be printed all in a single nozzle, significantly improving fabrication efficiency.

In this work, the UGI is printed by mixing two polyurethane dispersions (PUDs) with drastically different moduli, which allows the elastic modulus of the gradient material system to span two orders of magnitude, ranging from 1 MPa (soft PUD) to over 100 MPa (stiff PUD), (**Figure 1b**). Upon stretching the UGI, the deformation concentrates in low modulus regions (red colour) (**Figure 1b**), which makes the high modulus region (green colour) a strain-isolation region, on which printing the functional electronics is beneficial. Our AMMP approach achieves a high resolution of approximately 500 nm along the deposition thickness (Z direction) and 20 μm in the X-Y plane (**Figures S1** and **S2**). For demonstration, a 3D gradient UGI is displayed where digital image correlation (DIC) reveals a 2.5% maximum principal Lagrange strain in stiff areas, while the whole 3D gradient UGI is under 100% equi-biaxial stretch, showcasing its superior strain isolation capability (**Figure 1b**). This method effectively bridges the gap between soft tissues and a wide variety of stiff materials commonly used in electronic/sensing devices (**Figure 1c**).

To validate the strain-isolation performance of our method, we first printed UGIs to assess the effect of stiffness gradients on strain and stress distributions under various stretch ratios. Five different UGI designs were printed, including non-gradient, 1D in-plane, 2D in-plane, 1D out-of-plane, and 3D gradients (**Figures 2a, 2d, 2g,** and **S3**). Through *in situ* modulation of the material composition, these designs lead to different site-specific elastic moduli distributions (**Figures 2j**, **2k,** and **S4**). The 1D and 2D in-plane gradient designs reduce local stress concentrations arising from the sharp contrast of elastic moduli between the stiff and soft components. The out-of-plane and 3D gradient (**Figure S4**) add more dimensions to control the gradient properties and release the local stress concentration. The non-gradient design serves as a reference here (**Figure S4**).

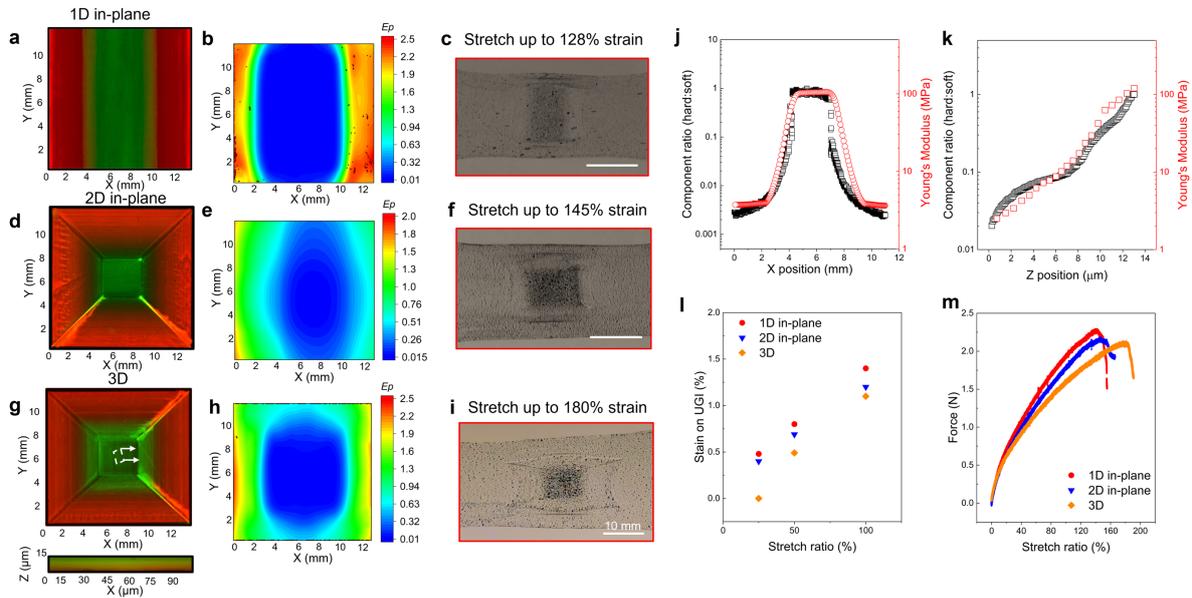

**Figure 2. Investigation of different UGI designs.** Gradient design, maximum principal Lagrange strain map ($E_p$) under 100% stretch ratio, and optical image under maximum stretch ratio for **a-c)**, 1D in-plane gradient, **d-f)**, 2D in-plane gradient, and **g-i),** 3D gradient. **j),** Composition and modulus distributions vs. X-positions for 1D and 2D in-plane gradient; **k),** Composition and modulus distributions vs. Z-positions for 3D gradient. **l),** Maximum principal Lagrange strain values on the center of the stiff region vs. external stretch for different UGI designs. **m),** tensile test results for different UGI designs.

The top surface strain maps obtained from DIC for samples under 100% global stretch show significant variations in strain distribution among different designs, particularly near the stiff regions and stiff/soft interfaces (**Figures 2b, 2e, 2h, S4, and S5**). In contrast, non-gradient designs demonstrate significant stress concentrations in the soft-stiff interfaces, and a crack occurs early at around 48% stretch in these regions (**Figures S4** and **S5**). Conversely, UGI effectively mitigates strain near the stiff areas by shifting the peak strain from these interfaces to the gradient regions, resulting in maximum principal Lagrange strains as low as 1.4%, 1.2%, and 1.1% for 1D in-plane, 2D in-plane, and 3D gradients in the stiff areas, respectively (**Figure 2l**). **Figure S5** shows excellent strain isolation for the 3D UGI where the strain in the stiff region is as small as 2.7% while the stretch ratio approaches 180%.

**Figures 2m** and **S7** display force-displacement curves for the different designs. Samples with the non-gradient designs fail at around 130% stretch ratio, while 3D gradient UGI breaks at an almost 1.5-fold higher stretch ratio. Additionally, the initial failure points differ across designs. Non-gradient and 1D out-of-plane samples typically fail at the stiff-soft interface, while UGIs such as 1D in-plane, 2D in-plane, and 3D gradients exhibit failure within the soft region under high strains (**Figures 2c, 2f, 2i, S4, S5,** and **S6**) due to the material's inherent strain limit. A Finite Element Analysis (FEA) model was created to analyse various failure modes in different designs and gain insights into local strains and stresses (**Figures S8** and **S9**). Non-gradient and 1D out-of-plane designs tend to fail prematurely due to stress and strain concentrations at the interface between stiff and soft regions, where there is a sharp change in modulus. In other designs, strain concentration occurs in the soft region and stress concentration in areas where a gradual transition from soft to stiff enhances failure resistance. Since stiff materials have higher failure stress, failure is less likely compared to non-gradient and 1D out-of-plane cases. This result underscores the crucial role of stress distribution in crack initiation and illustrates the significance of gradient design that facilitates more uniform stress distribution.

### 3. Versatility of UGI to enbale strain-insensitive stretcable materials and devices

To demonstrate the versatility and universal applications of the UGI, we printed conductive polymer films of PEDOT: PSS, 0D gold nanoparticle, 1D silver nanowires (AgNWs), and 2D MXene on the 3D UGI and subjected them up to 100% stretch ratio (see **Table S3** and **S4** for the detailed fabrication strategies). The printed Au film was sintered using a photonic flash sintering process detailed in **Figure S10**. The conductive films printed on homogeneous soft PU substrates were also tested for comparison. As shown in **Figures 3a-d,** all the conductive films on UGI show no visible cracks at a 100% stretch ratio, with the relative resistance change (RRC) ($\Delta R/R_0$) < 4.2%, whereas those on homogeneous substrates showing significant RRC on increases and microcracks leading to electrical failure. These results highlight the superior strain isolation properties of the UGI design compared to state-of-the-art stretchable electronics (**Figure 3e**)[9,27,31–41].

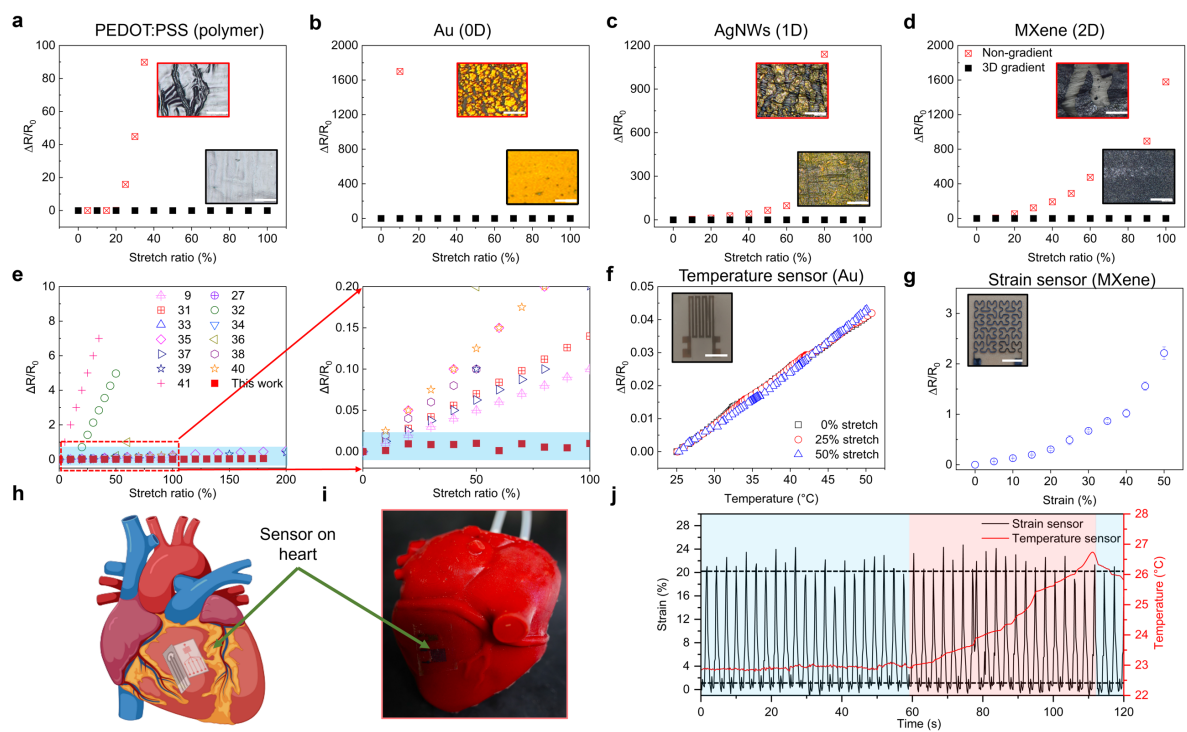

**Figure 3. Demonstration of the strain-insensitive effect of various conductive materials printed on the UGI.** Relative resistance change (RRC) vs. stretch ratio for **a)** PEDOT: PSS, **b)** Au, **c)** AgNWs, and **d)** MXene. Scale bar, 100 μm. **e)** Comparison of RRC in this work with previously reported stretchable electronics[9,27,31–41]. **f)** Temperature sensor printed on the UGI, and **g)** strain sensor printed on soft polyurethane. Scale bar, 4 mm. **h)** Schematic and **i)** optical image of multimodal temperature and strain sensor on an artificial heart model. **j)** Real-time strain and temperature sensing during the periodic diastole and systole when the heart model is under stable temperature followed by a transient heating.

Next, we explored the printing of electronic devices and sensors on the UGI. A stretchable Au-based resistive temperature sensor is printed on the 3D UGI (**Figure S12**), demonstrating a sensitivity of $1.6 \times 10^{-3}$ °C$^{-1}$ across a 25 to 50 °C range, with stable sensitivity during stretching (**Figure 3f**). Tools for cardiac physiological mapping are indispensable for the clinical identification and mechanistic understanding of excitation–contraction coupling, metabolic dysfunction, arrhythmia, and other conditions. Employing an artificial heart model, we evaluated the real-time performance of our sensors under dynamic conditions. To monitor the heart beating during diastole and systole, an MXene-based strain sensor was printed on the homogeneous soft substrate (**Figure 3g**). The strain and temperature sensors were then affixed to an artificial heart (**Figures 3h** and **3i**), which pulsates every two seconds. Our measurements show that the strain sensor can catch the periodic diastole and systole process (**Figure 3j**). Concurrently, the temperature sensor records temperature changes over time, maintaining high stability under repeated strain cycles (< 1.15% resistance change during consistent temperature conditions). These results indicate that the UGI provides a robust platform to monitor real-time signals in a dynamic environment.

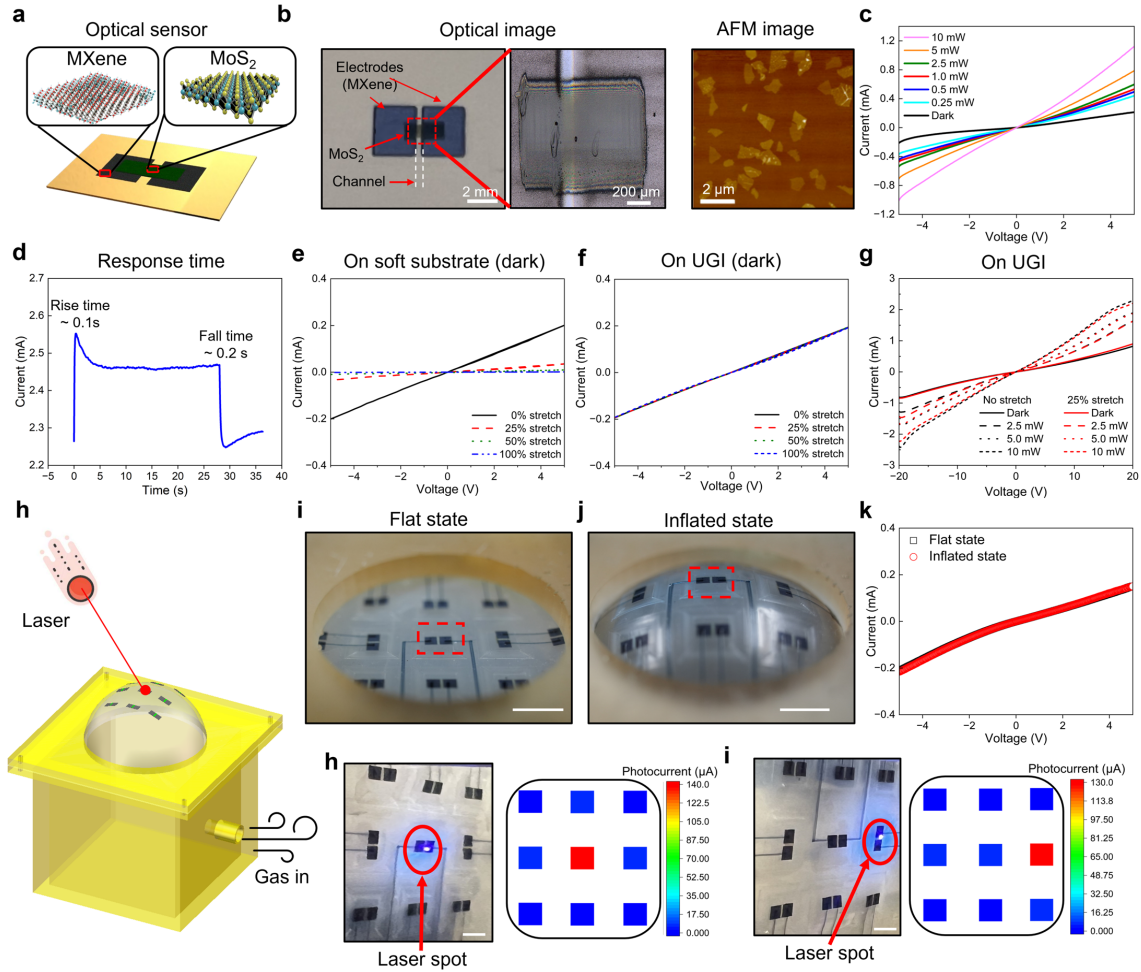

**Figure 4. Stretchable photodetectors. a)** Schematic, **b)** optical images and AFM[42] showing a printed photodetector on the UGI. **c)** Photocurrent vs. bias voltage under different illumination powers. **d)** Time-dependent photocurrent at an applied DC bias of 5V. **e-f)** Dark current vs. bias voltage under different stretch ratio on **e,** soft polyurethane and **f)** UGI. **g)** Photocurrent vs. bias voltage under different illumination powers when the UGI is under no stretch and 25% stretch **h)** Schematic of a deformable 3D photodetector array. **i-j)** Optical images of the 3D photodetector under flat and inflated state. **k)** Dark current vs. bias voltage during flat and inflated state. **l-m)** The photocurrent mapping under upright illumination and tilted-angle illumination. The radius of the hemisphere is 30 mm. Scale bar, 5 mm.

In addition to the stretchable resistance-based sensor, the UGI is employed to develop more sophisticated stretchable optoelectronic devices utilizing printed semiconducting 2D $MoS_2$, which transitions from an indirect to a direct bandgap in a monolayer regime and thus is ideal for optoelectronic applications in the atomically thin limit[43]. As illustrated in **Figures 4a** and **4b**, the $MoS_2$ inks were printed as the channel material between the two printed MXene electrodes (**Figure S14**), exhibiting a power-law relationship ($R \propto P^\gamma$, $\gamma = -0.06$), resulting in higher responsivities ($R$) at lower powers ($P$) (**Figure 4c** and **Figure S15**). At low laser intensity (0.15 W cm$^{-2}$), the responsivity for the devices reached 0.62 A W$^{-1}$, which is among the highest of previously reported all-printed photodetectors in the visible light range. Further details on the responsivity calculation are provided in **Figure S15**. Photodetector response times, with rise and fall constants of approximately 0.1 s and 0.2 s, respectively, reflect the rapid generation and slower recombination of charge carriers facilitated by deep trap states (**Figure 4d**)[44]. A trade-off between response time and MoS$_2$ layer thickness is observed (**Figure S17**). Our all-printed photodetectors exhibit relatively fast response times compared to some literature using CVD-grown and mechanically exfoliated MoS$_2$ photodetectors (**Figure S16**)[45–47]. The slower response times of CVD-grown and mechanically MoS$_2$ photodetectors, linked to interface traps at the MoS$_2$-SiO$_2$ interface, contrast with faster times on PU substrate due to reduced trapping[45,48].

To assess the performance of photodetectors on both 3D gradient and soft interfaces, we recorded the dark current under various stretch ratios (**Figures 4e** and **4f**). Even under 100% stretch, the dark current shows <2.9% change on the 3D gradient interface, whereas over 16700% decrease of dark current occurs on the soft substrate. Performance on 3D gradient interfaces under various laser powers and strains demonstrated <2.6% signal drift at 25% stretch (**Figure 4g**), highlighting the robust strain isolation of the UGI design, crucial for applications in dynamically changing environments like tunable electronic eye cameras.

An array of nine photodetectors was printed and mounted on an inflatable hemispherical structure subjected to variable 2D strains to demonstrate advanced 3D sensing systems. (**Figure 4h-j** and **Figure S17**). The photodetector array exhibits remarkable strain isolation capabilities and <1.9% change of dark current under maximum in-plane strains of 57% in each direction (**Figure 4k**). Finite element simulations (**Figure S18**) revealed that the maximum principal strain in the photodetectors on the UGI is minimal, indicating the robustness of this design.

The 3D stretchable photodetector array can simultaneously detect both the light intensity and discern the position of incident light. When laser light (405 nm wavelength) directly strikes the center of the photodetector array, the central detector exhibits a dominant photocurrent of 139.6 μA, with the surrounding eight detectors showing negligible photocurrents (**Figure 4l**). However, for oblique incident light to the right sensor, the most intense response emerges in the laser-light-focused photosensor (right sensor) (**Figures 4m** and **S19**). These findings pave the way for the integration of stretchable photodetectors in versatile 3D sensing applications.

Envisioning future advancements, combining these sensors with 3D-printed laser, visible, and infrared detectors could revolutionize bionic eye development, offering comprehensive all-angle and all-wavelength visual capabilities.

## 4. All printed strain-insenstive on-skin sensors for continuous health monitoring

The UGI serves as a versatile platform for printing multimodal sensors for on-skin health monitoring. Here we demonstrate harmonious integration of temperature and blood oxygen saturation sensors on the stiff region of the UGI and a pulse sensor on the soft region to detect pulse-induced strain changes, creating a stretchable system integrated seamlessly with the skin (**Figures 5a**, **5b**, and **S20**).

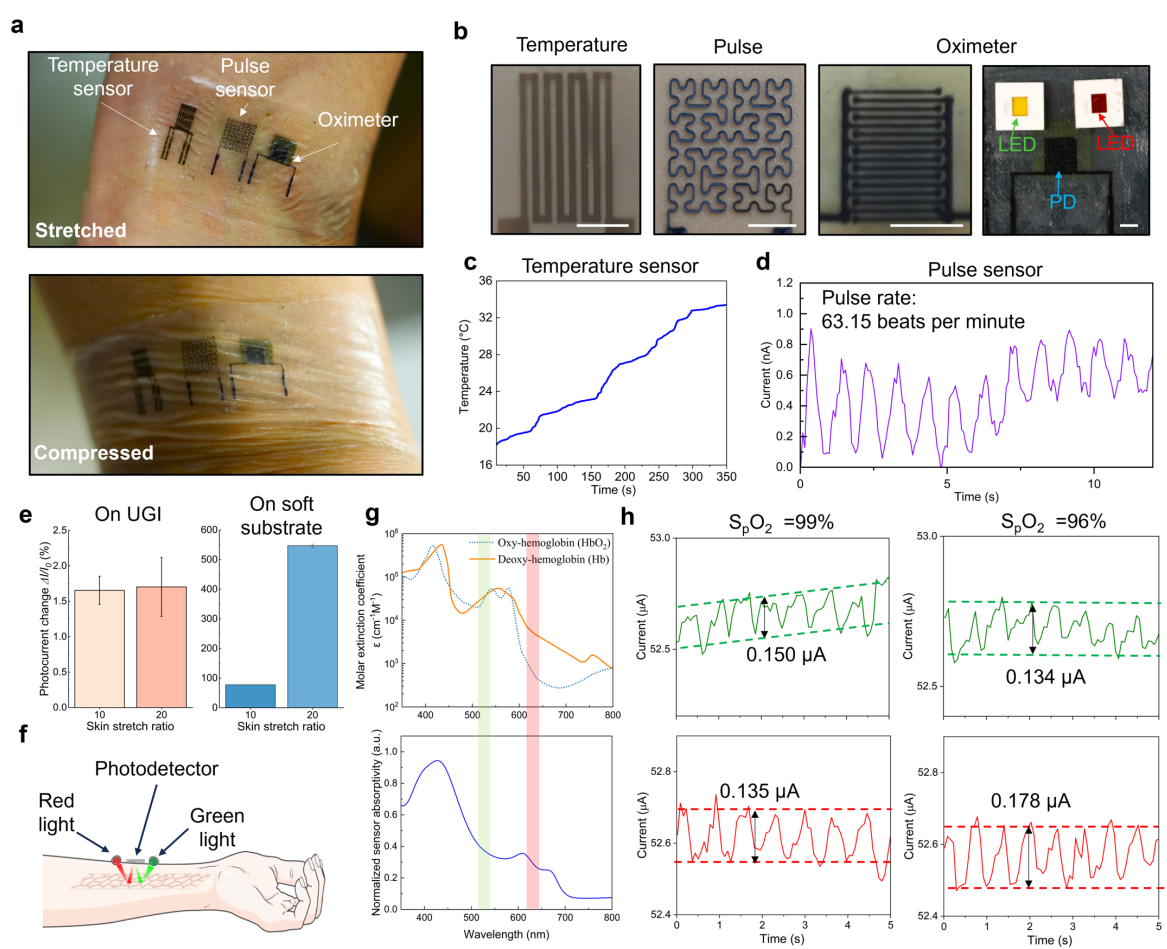

**Figure 5. All printed wearable multimodal sensors for health monitoring. a)** Optical image of the multimodal sensor. **b)** Optical images of the temperature sensor**,** pulse sensor, and oximeter consisting of a printed photodetector (PD) and a green LED and a red LED. Scale bar, 2 mm. Characteristics of the **c)** temperature sensor and **d)** pulse sensor. **e)** Motion artifacts for oximeter printed on UGI and soft substrate. **f)** Schematic of the oximeter in reflectance mode. **g)** Molar extinction coefficient of oxygenated (orange solid line) and deoxygenated (blue dashed line) haemoglobin in arterial blood, and absorptivity of oximeter vs. Wavelength. **h)** Output signal from oximeter with 99% and 96% oxygenation of blood. The green and red lines represent the signals when the green and red LEDs are operated.

The temperature sensor accurately tracks skin temperature fluctuations, capturing changes from 18.2°C to 33.4°C as the participant moves from a cold to a warm environment (**Figure 5c**). The pulse sensor using printed MXene patterns dynamically records the electrical current variations that correspond with the heartbeat, enabling precise heart rate monitoring by counting peaks over a minute (**Figure 5d**). Pulse oximetry, a widely utilized non-invasive method, evaluates tissue oxygen saturation by optically determining the relative concentrations of oxyhemoglobin ($HbO_2$) and deoxyhemoglobin (Hb) in blood using photoplethysmography (PPG). Our $MoS_2$-based photodetector is integrated with green and red LED lights (**Figure 5f**) to create a stretchable pulse oximeter that conforms to the skin. In this setup, reflected light is modulated by the dynamic volumes of arterial blood, affecting light absorption detected by the photodetector. This modulation, correlating with the heart's systolic and diastolic phases, produces a pulsatile optical signal that is analysed to determine the blood oxygenation levels (**Figure S21**). $HbO_2$ and Hb exhibit different absorption rates of oxygenated and deoxygenated hemoglobin at red and green wavelengths (**Figure 5g**). Blood oxygen saturation ($SpO_2$) can be calculated by determining the light absorption ratio at these two wavelengths (**Figure S21**). **Figure 5h** shows the corresponding signals when $SpO_2$ decreases from 99% to 96% during a breath hold test.

Motion artifacts pose significant challenges in wearable electronics for health monitoring, leading to potential misinterpretations and severe health implications. To demonstrate the superior immunity to strain/motion artifacts of printed UGI, photodetectors were printed on both UGI and soft substrates respectively, and then tested on skin with induced strain deformations while recording the signals (**Figure S22**). The oximeter on the UGI showed neglectable signal deviation (<1.7% change of measured photocurrent under approximately 20% strain), as depicted in **Figure 5e,** demonstrating by far the best performances in reducing motion artifacts among previously reported wearable devices (**Table S3**). In contrast, the oximeter on the soft substrate experienced over >547% signal variance (**Figure 5e**), underscoring the UGI's superior capability for stable biological signal capture amidst skin deformations.

## 5. Conclusion

In summary, the printed universal gradient interfaces (UGIs) with submicron resolution and 3D modulus distributions provide unprecedented strain isolation and a versatile platform for realizing strain-insensitive stretchable devices with over 180% stretchability. The versatility of aerosol-based multi-materials printing (AMMP) method allows for integrating a diverse range of structural and functional materials into stretchable systems with precise control over global device structures and local properties with unparalleled spatial resolution. This UGI platform enables multifunctional stretchable devices for on-skin sensing of temperature, light, and blood oxygen, which allows truly motion artifact-free health monitoring. This innovation paves the way for advanced wearable and implantable devices for real-time health monitoring and therapeutics as well as for soft and hybrid robotics with integrated soft-stiff gradient interfaces that are essential in natural musculoskeletal structures.

## 6. Experimental Section

*General ink formulation*: Commercially available gold ink (10 wt.% in Xylene) was purchased from UT Dots. AgNWs (1 wt.% in isopropanol) ink was purchased from Nanostructured & Amorphous Materials. PEDOT: PSS (0.8 wt.% in water) were purchased from Sigma-Aldrich. Soft (Baymedix CD 102) and hard PUD (U 9380) inks were obtained from Covestro and Alberdingk Boley, respectively. Depending on the ink type, a small amount of isopropyl alcohol (IPA) was often added as a defoamer to suppress foam formation during ultrasonic atomizing. To avoid aggregation and ensure uniform dispersion, inks were sonicated for 15 min (Fisher Scientific sonicate bath, 9.5L). Typical ink composition can be found in **Table S4**. The concentration of nanomaterial was determined by the drying weight method.

For the preparation of the $MoS_2$ ink, a two-electrode electrochemical cell was used for electrochemical intercalation. A thin piece of cleaved $MoS_2$ crystal was clamped by a copper alligator clip as the cathode, and a graphite rod was placed as the anode respectively. Tetraheptylammonium bromide (THAB)/acetonitrile (5 mg mL$^{-1}$) served as the electrolyte. The electrochemical intercalation was conducted under a voltage of 7.5 V for 1 hour. Afterward, the as-intercalated $MoS_2$ was rinsed with DMF and sonicated in PVP/DMF (0.2 mol L$^{-1}$, Mw=40, 000) for 30 min. The dispersion was centrifuged and washed twice with IPA to remove excessive PVP. Then, the dispersion in IPA was centrifuged at 3000 rpm for 5 min, and the precipitates were discarded. The final dispersion of exfoliated $MoS_2$ ink in IPA was ready for further printing.

The synthesis of $Ti_3C_2T_x$ MXene ink involved selectively etching 1 g of $Ti_3AlC_2$ with an acidic solution of 9 mL deionized water, 3 mL hydrofluoric acid, and 18 mL hydrochloric acid. This mixture was stirred in a high-density polyethylene container on a heated stir plate for 24 hours. Post-etching, the solution was centrifuged and washed to a neutral pH, followed by delamination using anhydrous lithium chloride. The resulting slurry was further washed, centrifuged, and concentrated into ink, which was then dried in a vacuum oven at 100 °C, resulting in a free-standing film. Prior to shipping, the ink is treated with argon and sealed in aluminum foil for overnight delivery.

*Fabrication of the strain-isolating substrate and sensors*: The motion control graphical user interface (GUI) controls the printer's x, y, and z motion stages. Real-time position and velocity were tracked for x, y, and z during the targeted move and jogging operations. Both aerosol ink flow rates were actively controlled concerning stage position to achieve gradient and voxelated films. For all high-resolution printing, the inks were primed under sonication for 30 min and then atomized via ultrasonication before being transferred to the printhead. A sheath flow focused the aerosolized ink stream to achieve high printing resolution. Before printing, gradient or homogeneous substrates were treated with plasma to improve surface hydrophilicity. During the AMMP process, a heating stage evaporated the ink solvents to minimize undesired drying effects. Depending on the type of combinatorial materials, additional flash sintering was adopted to achieve the desired microstructures and properties.

All designs of substrates and sensors were printed using AMMP. First, the substrates were designed in AutoCAD, and then the printing paths were automatically generated. To measure the strain distribution, we prepared all specimens with a length of 31 mm, a width of 21 mm, and a gradient composition. The process parameters, including atomizer settings, ink qualities, carrier and sheath gas flow conditions, printer head as well as nozzle geometry, the distance between the nozzle tip and substrate, substrate temperature, etc., are summarized in **Table S5** and **Table S6**.

*Digital image correlation (DIC)*: The 2D DIC method was used to characterize the strain distribution and probe Young's modulus distribution of all 5 cases of printed substrates (1D non-gradient, 1D in-plane gradient, 1D out-of-plane gradient, 2D in-plane gradient, and 3D gradient). The force-displacement relation was recorded by uniaxial stretching of the specimen up to failure at a rate of 0.5% (of the initial length) per second via an Instron universal machine (model 5944) with a 50 N load cell. The specimen was mounted in a pair of pneumatic tensile grips, leaving a gauge length of 21 mm. To measure the strain distribution by the DIC method, an ink (Koh-I-Noor Rapidraw) was sprayed with an airbrush (Badger, no. 150) to generate high-quality speckle patterns on the specimen. A whiteboard was used as a background, and a white LED light was shone on the sample during testing to enhance the optical contrast. A Canon ESO 6D DSLR camera recorded changes in the speckle patterns with a Canon 100 mm F/2.8L macro lens at roughly every 2% strain. The resolution of each image was around 15 μm per pixel, and data was recorded for every four-pixel length. Images were analyzed by Ncorr40, an open-source 2D DIC MATLAB software, to obtain the Eulerian strain distributions of the middle region of 11 mm (L) × 11 mm (W). Based on the strain distribution, we further calculated the modulus distribution, the details of which were elaborated in our previous work[49].

*Finite element analysis (FEA)*: The design can be applied to substrates in wearable devices. We assumed the sample could be intimately attached to an arbitrary surface, such as the wrist, and stretched when rotating and bending. The sample can deform with the skin without failure. We utilized the commercial finite element software ABAQUS to simulate a gradient substrate's stress and strain distributions under uniaxial and biaxial tension. We built a 3D model with various modulus distributions designed along the x, y, and z directions. The thickness of the uniform soft substrate is 10 μm. On the top of the uniform substrate is a gradient square part with a length of 11 mm and a thickness of 1 μm (**Figure S8)**. The stress and strain contours in the x-y plane are shown in **Figures S8** and **S9**. By symmetry, only a quarter of the specimen is modeled separately with XSYMM and YSYMM boundary conditions at the left and bottom. We assigned an ABAQUS built-in hyperelastic model with the neo-Hookean form to the whole model. We set the shear moduli for the soft and hard PUD as 0.75 MPa to 75 MPa, respectively, and the compressibility parameter to be D = 0 to enforce incompressibility. The sample was stretched in the X direction up to 100% strain for the uniaxial tension case and simultaneously in the X and Y directions up to 100% strain for the biaxial tension case. We applied element type C3D8H and nonlinear static analysis.

*Nanoindentation*: Nanoindentation experiments were conducted using a nanoindenter (Hysitron TriboIndenter, Bruker, Billerica, MA), applying a force of 70 μN to determine the elastic modulus and hardness of the sample material.

*White light profilometer*: Sample thickness was measured using a white light profilometer (Filmetrics, San Diego, CA) equipped with a 5X magnification objective, which employs chromatic aberration to accurately gauge the vertical displacement between the highest and lowest points of the sample's surface.

*Fluorescent imaging*: Two PUD inks were mixed with different fluorescent dyes and imaged to investigate the gradient composition of printed substrates. Specifically, a red colour dye and a green colour dye were added to the PUDs to visualize the composition change process better. The fluorescent dyes were purchased from GLO EFFEX with red (UVT-RD-1OZ) and green (UVT-GR-1OZ). The fluorescent images were captured with a Keyence BZ-X810 Microscope. RGB analyses quantified the intensity of the two colours on the printed gradient polyurethane films, where the RGB colour profiles were obtained by using the ImageJ RGB-Profiler plugin (National Institutes of Health, USA).

*Resistance analysis*: A custom-built test setup was used to measure the resistance of electronics under different strains. The straining mechanism consisted of a stepper motor for driving the timing belt, a microcontroller unit to control the precise movement of the timing belt, and a base connected to the timing belt that holds the tested substrate to be stretched. The resistance of the strain gauge is measured using the DAQ system and LabVIEW. The data was continuously monitored and recorded on a computer.

*Flash sintering*: Flash sintering was performed using a Sinteron 2100 system (Xenon Corp., USA) with a 107 mm xenon spiral lamp. The S-2100 system was configured for maximum pulse durations of 3 ms, with the sintering carried out in an ambient environment. The S-2100 produced pulse energy (single) ranging from 30 to 2850 J.

*Photodetector characterization*: Photoresponse measurements were conducted using a Keithley 2636A source meter under ambient conditions. A 405 nm laser (Crystal Laser) served as the light source, with power adjustments made via the diode current. Calibration was achieved using an energy meter (PM100D, Thor Labs). Spectrally resolved photocurrents were recorded across wavelengths from 360 to 800 nm at a constant power of 2.5 mW using a supercontinuum broadband laser (SuperK FIANIUM-NKT photonics) and bandwidth tunable filter (SuperK VARIA-NKT photonics). Response times were determined by dynamically modulating the laser with a shutter, to control the periods of light exposure. The driving voltage of the red and green LEDs was set at 3 V. At the same time, the oxygen saturation was measured by a commercial pulse oximeter (CMS50NA, Contact Medical Systems Co., Ltd.). The oxygen saturation ($SpO_2$) can be expressed as a function of transmitted light ratio ($R_{OS}$) of green and red LEDs. We calculated the green and red PPG signal amplitudes and then calculated the $R_{OS}$. An informed written consent from all human participants was obtained prior to the research.

**Supporting Information**

Supporting Information is available from the Wiley Online Library or from the author.

**Acknowledgements**
Y.Z. acknowledges funding support from the National Science Foundation (award no. CMMI-1747685) and the US Department of Energy (award no. DE-EE0009103). X.D. acknowledges the support from the Office of Naval Research (award no. N00014-22-1-2631). The authors acknowledge the use and support of the Notre Dame Materials Characterization Facility and Analytical Science and Engineering Core Facility. This work used the computational and storage services associated with the Hoffman2 Shared Cluster provided by Institute for Digital Research and Education's Research Technology Group at the University of California, Los Angeles. No formal approval for the experiments related to wearable skin-technologies involving human volunteers was required by the authors' institute.

Received: ((will be filled in by the editorial staff))

Revised: ((will be filled in by the editorial staff))

Published online: ((will be filled in by the editorial staff))

References

[1] J. A. Rogers, T. Someya, Y. Huang, *Science (1979)* **2010**, *327*, 1603.
[2] W. Wang, S. Wang, R. Rastak, Y. Ochiai, S. Niu, Y. Jiang, P. K. Arunachala, Y. Zheng, J. Xu, N. Matsuhisa, X. Yan, S. K. Kwon, M. Miyakawa, Z. Zhang, R. Ning, A. M. Foudeh, Y. Yun, C. Linder, J. B. H. Tok, Z. Bao, *Nat Electron* **2021**, *4*, 143.
[3] S. Wang, J. Xu, W. Wang, G. J. N. Wang, R. Rastak, F. Molina-Lopez, J. W. Chung, S. Niu, V. R. Feig, J. Lopez, T. Lei, S. K. Kwon, Y. Kim, A. M. Foudeh, A. Ehrlich, A. Gasperini, Y. Yun, B. Murmann, J. B. H. Tok, Z. Bao, *Nature* **2018**, *555*, 83.
[4] Y. Dai, H. Hu, M. Wang, J. Xu, S. Wang, *Nat Electron* **2021**, *4*, 17.
[5] F. Ershad, A. Thukral, J. Yue, P. Comeaux, Y. Lu, H. Shim, K. Sim, N. I. Kim, Z. Rao, R. Guevara, L. Contreras, F. Pan, Y. Zhang, Y. S. Guan, P. Yang, X. Wang, P. Wang, X. Wu, C. Yu, *Nat Commun* **2020**, *11*, 1.
[6] D. Kireev, S. K. Ameri, A. Nederveld, J. Kampfe, H. Jang, N. Lu, D. Akinwande, *Nat Protoc* **2021**, *16*, 2395.
[7] Y. Hui, Y. Yao, Q. Qian, J. Luo, H. Chen, Z. Qiao, Y. Yu, L. Tao, N. Zhou, *Nat Electron* **2022**, *5*, 893.
[8] T. Sakorikar, N. Mihaliak, F. Krisnadi, J. Ma, T. Il Kim, M. Kong, O. Awartani, M. D. Dickey, *Chem Rev* **2024**, *124*, 860.
[9] Y. Zhao, B. Wang, J. Tan, H. Yin, R. Huang, J. Zhu, S. Lin, Y. Zhou, D. Jelinek, Z. Sun, K. Youssef, L. Voisin, A. Horrillo, K. Zhang, B. M. Wu, H. A. Coller, D. C. Lu, Q. Pei, S. Emaminejad, *Science (1979)* **2022**, *378*, 1222.
[10] Q. Shen, M. Jiang, R. Wang, K. Song, M. Hou Vong, W. Jung, F. Krisnadi, R. Kan, F. Zheng, B. Fu, P. Tao, C. Song, G. Weng, B. Peng, J. Wang, W. Shang, M. D. Dickey, T. Deng, *Science (1979)* **2023**, *379*, 488.
[11] Y. S. Guan, F. Ershad, Z. Rao, Z. Ke, E. C. da Costa, Q. Xiang, Y. Lu, X. Wang, J. Mei, P. Vanderslice, C. Hochman-Mendez, C. Yu, *Nat Electron* **2022**, *5*, 881.
[12] Y. Luo, M. R. Abidian, J. H. Ahn, D. Akinwande, A. M. Andrews, M. Antonietti, Z. Bao, M. Berggren, C. A. Berkey, C. J. Bettinger, J. Chen, P. Chen, W. Cheng, X. Cheng, S. J. Choi, A. Chortos, C. Dagdeviren, R. H. Dauskardt, C. A. Di, M. D. Dickey, X. Duan, A. Facchetti, Z. Fan, Y. Fang, J. Feng, X. Feng, H. Gao, W. Gao, X. Gong, C. F. Guo, X. Guo, M. C. Hartel, Z. He, J. S. Ho, Y. Hu, Q. Huang, Y. Huang, F. Huo, M. M. Hussain, A. Javey, U. Jeong, C. Jiang, X. Jiang, J. Kang, D. Karnaushenko, A. Khademhosseini, D. H. Kim, I. D. Kim, D. Kireev, L. Kong, C. Lee, N. E. Lee, P. S. Lee, T. W. Lee, F. Li, J. Li, C. Liang, C. T. Lim, Y. Lin, D. J. Lipomi, J. Liu, K. Liu, N. Liu, R. Liu, Y. Liu, Y. Liu, Z. Liu, Z. Liu, X. J. Loh, N. Lu, Z. Lv, S. Magdassi, G. G. Malliaras, N. Matsuhisa, A. Nathan, S. Niu, J. Pan, C. Pang, Q. Pei, H. Peng, D. Qi, H. Ren, J. A. Rogers, A. Rowe, O. G. Schmidt, T. Sekitani, D. G. Seo, G. Shen, X. Sheng, Q. Shi, T. Someya, Y. Song, E. Stavrinidou, M. Su, X. Sun, K. Takei, X. M. Tao, B. C. K. Tee, A. V. Y. Thean, T. Q. Trung, C. Wan, H. Wang, J. Wang, M. Wang, S. Wang, T. Wang, Z. L. Wang, P. S. Weiss, H. Wen, S. Xu, T. Xu, H. Yan, X. Yan, H. Yang, L. Yang, S. Yang, L. Yin, C. Yu, G. Yu, J. Yu, S. H. Yu, X. Yu, E. Zamburg, H. Zhang, X. Zhang, X. Zhang, X. Zhang, Y. Zhang, Y. Zhang, S. Zhao, X. Zhao, Y. Zheng, Y. Q. Zheng, Z. Zheng, T. Zhou, B. Zhu, M. Zhu, R. Zhu, Y. Zhu, Y. Zhu, G. Zou, X. Chen, *ACS Nano* **2023**, *17*, 5211.
[13] D. G. Mackanic, T. H. Chang, Z. Huang, Y. Cui, Z. Bao, *Chem Soc Rev* **2020**, *49*, 4466.
[14] K. Keum, J. W. Kim, S. Y. Hong, J. G. Son, S. S. Lee, J. S. Ha, *Advanced Materials* **2020**, *32*, 1.

[15] Z. Huang, Y. Hao, Y. Li, H. Hu, C. Wang, A. Nomoto, T. Pan, Y. Gu, Y. Chen, T. Zhang, W. Li, Y. Lei, N. H. Kim, C. Wang, L. Zhang, J. W. Ward, A. Maralani, X. Li, M. F. Durstock, A. Pisano, Y. Lin, S. Xu, *Nat Electron* **2018**, *1*, 473.

[16] M. Baumgartner, F. Hartmann, M. Drack, D. Preninger, D. Wirthl, R. Gerstmayr, L. Lehner, G. Mao, R. Pruckner, S. Demchyshyn, L. Reiter, M. Strobel, T. Stockinger, D. Schiller, S. Kimeswenger, F. Greibich, G. Buchberger, E. Bradt, S. Hild, S. Bauer, M. Kaltenbrunner, *Nat Mater* **2020**, *19*, 1102.

[17] H. Cho, B. Lee, D. Jang, J. Yoon, S. Chung, Y. Hong, *Mater Horiz* **2022**, *9*, 2053.

[18] J. Lee, J. Wu, M. Shi, J. Yoon, S. Il Park, M. Li, Z. Liu, Y. Huang, J. A. Rogers, *Advanced Materials* **2011**, *23*, 986.

[19] Y. K. Lee, K. I. Jang, Y. Ma, A. Koh, H. Chen, H. N. Jung, Y. Kim, J. W. Kwak, L. Wang, Y. Xue, Y. Yang, W. Tian, Y. Jiang, Y. Zhang, X. Feng, Y. Huang, J. A. Rogers, *Adv Funct Mater* **2017**, *27*, 1.

[20] J. C. Yang, S. Lee, B. S. Ma, J. Kim, M. Song, S. Y. Kim, D. W. Kim, T. S. Kim, S. Park, *Sci Adv* **2022**, *8*, DOI 10.1126/sciadv.abn3863.

[21] S. Xu, Y. Zhang, L. Jia, K. E. Mathewson, K. I. Jang, J. Kim, H. Fu, X. Huang, P. Chava, R. Wang, S. Bhole, L. Wang, Y. J. Na, Y. Guan, M. Flavin, Z. Han, Y. Huang, J. A. Rogers, *Science (1979)* **2014**, *344*, 70.

[22] M. Cai, S. Nie, Y. Du, C. Wang, J. Song, *ACS Appl Mater Interfaces* **2019**, *11*, 14340.

[23] H. Liu, M. H. and stretchable integrated electroni Li, S. Liu, P. Jia, X. Guo, S. Feng, T. J. Lu, H. Yang, F. Li, F. Xu, *Mater Horiz* **2020**, 203.

[24] Y. Li, N. Li, W. Liu, A. Prominski, S. Kang, Y. Dai, Y. Liu, H. Hu, S. Wai, S. Dai, Z. Cheng, Q. Su, P. Cheng, C. Wei, L. Jin, J. A. Hubbell, B. Tian, S. Wang, *Nat Commun* **2023**, *14*, DOI 10.1038/s41467-023-40191-3.

[25] J. Byun, E. Oh, B. Lee, S. Kim, S. Lee, Y. Hong, *Adv Funct Mater* **2017**, *27*, 1.

[26] Y. Ma, M. Pharr, L. Wang, J. Kim, Y. Liu, Y. Xue, R. Ning, X. Wang, H. U. Chung, X. Feng, J. A. Rogers, Y. Huang, *Small* **2017**, *13*, 1.

[27] R. Libanori, R. M. Erb, A. Reiser, H. Le Ferrand, M. J. Süess, R. Spolenak, A. R. Studart, *Nat Commun* **2012**, *3*, 1.

[28] C. T. Forte, S. M. Montgomery, L. Yue, C. M. Hamel, H. J. Qi, *J Appl Mech* **2023**, *90*, 1.

[29] D. Kokkinis, F. Bouville, A. R. Studart, *Advanced Materials* **2018**, *30*, 1.

[30] M. Kim, S. Hong, J. J. Park, Y. Jung, S. H. Choi, C. Cho, I. Ha, P. Won, C. Majidi, S. H. Ko, *Advanced Materials* **2024**, *2313344*, 1.

[31] K. H. Kim, M. Vural, M. F. Islam, *Advanced Materials* **2011**, *23*, 2865.

[32] A. Miyamoto, S. Lee, N. F. Cooray, S. Lee, M. Mori, N. Matsuhisa, H. Jin, L. Yoda, T. Yokota, A. Itoh, M. Sekino, H. Kawasaki, T. Ebihara, M. Amagai, T. Someya, *Nat Nanotechnol* **2017**, *12*, 907.

[33] M. Shin, J. H. Song, G. H. Lim, B. Lim, J. J. Park, U. Jeong, *Advanced Materials* **2014**, *26*, 3706.

[34] G. D. Moon, G. H. Lim, J. H. Song, M. Shin, T. Yu, B. Lim, U. Jeong, *Advanced Materials* **2013**, *25*, 2707.

[35] P. Lee, J. Lee, H. Lee, J. Yeo, S. Hong, K. H. Nam, D. Lee, S. S. Lee, S. H. Ko, *Advanced Materials* **2012**, *24*, 3326.

[36] D. C. Hyun, M. Park, C. Park, B. Kim, Y. Xia, J. H. Hur, J. M. Kim, J. J. Park, U. Jeong, *Advanced Materials* **2011**, *23*, 2946.

[37] V. R. Feig, H. Tran, M. Lee, Z. Bao, *Nat Commun* **2018**, *9*, 1.

[38] J. Park, S. Choi, A. H. Janardhan, S. Y. Lee, S. Raut, J. Soares, K. Shin, S. Yang, C. Lee, K. W. Kang, H. R. Cho, S. J. Kim, P. Seo, W. Hyun, S. Jung, H. J. Lee, N. Lee, S. H. Choi, M. Sacks, N. Lu, M. E. Josephson, T. Hyeon, D. H. Kim, H. J. Hwang, *Sci Transl Med* **2016**, *8*, 1.

[39] R. Ma, B. Kang, S. Cho, M. Choi, S. Baik, *ACS Nano* **2015**, *9*, 10876.
[40] S. Choi, J. Park, W. Hyun, J. Kim, J. Kim, Y. B. Lee, C. Song, H. J. Hwang, J. H. Kim, T. Hyeon, D. H. Kim, *ACS Nano* **2015**, *9*, 6626.
[41] T. Yang, Y. Zhong, D. Tao, X. Li, X. Zang, S. Lin, X. Jiang, Z. Li, H. Zhu, *2d Mater* **2017**, *4*, DOI 10.1088/2053-1583/aa78cc.
[42] Z. Lin, Y. Liu, U. Halim, M. Ding, Y. Liu, Y. Wang, C. Jia, P. Chen, X. Duan, C. Wang, F. Song, M. Li, C. Wan, Y. Huang, X. Duan, *Nature* **2018**, *562*, 254.
[43] L. Kuo, V. K. Sangwan, S. V. Rangnekar, T. C. Chu, D. Lam, Z. Zhu, L. J. Richter, R. Li, B. M. Szydłowska, J. R. Downing, B. J. Luijten, L. J. Lauhon, M. C. Hersam, *Advanced Materials* **2022**, *34*, 1.
[44] Y. Li, L. Li, S. Li, J. Sun, Y. Fang, T. Deng, *ACS Omega* **2022**, *7*, 13615.
[45] O. Lopez-Sanchez, D. Lembke, M. Kayci, A. Radenovic, A. Kis, *Nat Nanotechnol* **2013**, *8*, 497.
[46] W. Zhang, J. K. Huang, C. H. Chen, Y. H. Chang, Y. J. Cheng, L. J. Li, *Advanced Materials* **2013**, *25*, 3456.
[47] S. Li, X. Chen, F. Liu, Y. Chen, B. Liu, W. Deng, B. An, F. Chu, G. Zhang, S. Li, X. Li, Y. Zhang, *ACS Appl Mater Interfaces* **2019**, *11*, 11636.
[48] A. Di Bartolomeo, L. Genovese, T. Foller, F. Giubileo, G. Luongo, L. Croin, S. J. Liang, L. K. Ang, M. Schleberger, *Nanotechnology* **2017**, *28*, DOI 10.1088/1361-6528/aa6d98.
[49] M. Zeng, Y. Du, Q. Jiang, N. Kempf, C. Wei, M. V. Bimrose, A. N. M. Tanvir, H. Xu, J. Chen, D. J. Kirsch, J. Martin, B. C. Wyatt, T. Hayashi, M. Saeidi-Javash, H. Sakaue, B. Anasori, L. Jin, M. D. McMurtrey, Y. Zhang, *Nature* **2023**, *617*, 292.

# Supplementary Information

**A Printed Microscopic Universal Gradient Interface for Super Stretchable Strain-Insensitive Bioelectronics**

*Kaidong Song, Jingyuan Zhou, Chen Wei, Ashok Ponnuchamy, Md Omarsany Bappy, Yuxuan Liao, Qiang Jiang, Yipu Du, Connor J. Evans, Brian C. Wyatt, Thomas O' Sullivan, Ryan K. Roeder, Babak Anasori, Anthony J. Hoffman[4], Lihua Jin, Xiangfeng Duan∗ and Yanliang Zhang∗*

K. Song, M. O. Bappy, Y. Liao, Q. Jiang, Y. Du, C. Evans, R. Roeder, Y. Zhang
Department of Aerospace and Mechanical Engineering
University of Notre Dame
Notre Dame, IN 46556, USA
E-mail: yzhang45@nd.edu

J. Zhou, X. Duan
Chemistry and Biochemistry Department
University of California Los Angeles
Los Angeles, CA 90095, USA
E-mail: xduan@chem.ucla.edu

C. Wei, L. Jin
Department of Mechanical and Aerospace Engineering
University of California Los Angeles
Los Angeles, CA 90095, USA

A. Ponnuchamy, Thomas O' Sullivan, A. Hoffman
Department of Electrical Engineering
University of Notre Dame
Notre Dame, IN 46556, USA

B. Wyatt, B. Anasori
School of Materials Engineering
Purdue University
West Lafayette, IN 47907, USA

**Supplementary Note 1. XY-plane printing resolution**

Our aerosol-based multi-materials printing (AMMP) approach achieves a high resolution of approximately 20 μm in the X-Y plane. Here are the single filaments printed through soft PUD and hard PUD inks.

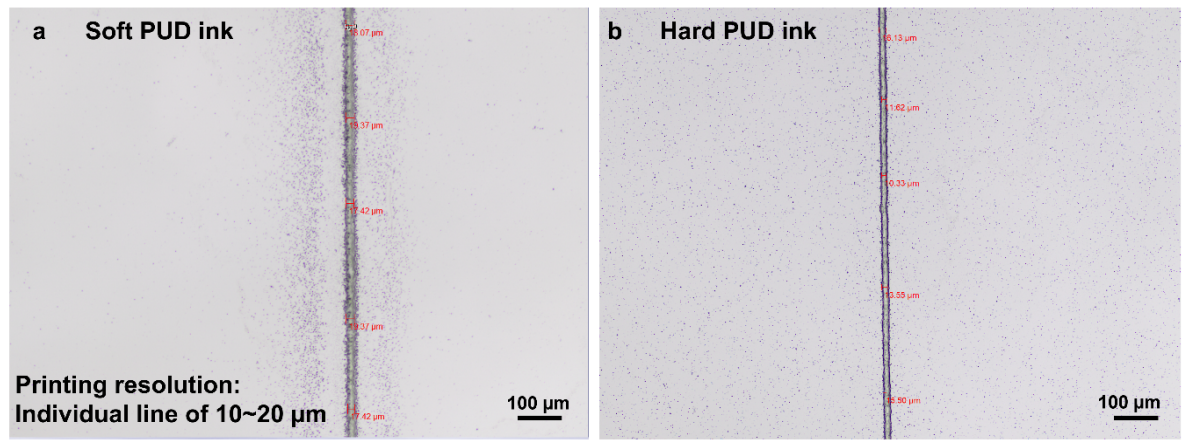

**Figure S1. XY-plane printing resolution.** Printing of an individual line showing the X-Y plane resolution as small as 10 μm for (a) soft PUD, and (b) hard PUD inks (Nozzle size: 30 gauge).

**Supplementary Note 2. Z-plane printing resolution**

**Figure S2a** shows the 3D scan of the printed 1-layer PUD film by using a white light profilometer (Filmetrics, Profilm3D). White light interferometry technology is used by this kind of profilometer to provide quantitative surface topological information. It is a nondestructive, non-contact type of measurement method offering good technology to measure film thickness. The film thickness shown in **Figure S2b** demonstrates the deposition resolution along Z direction as small as around 500 nm.

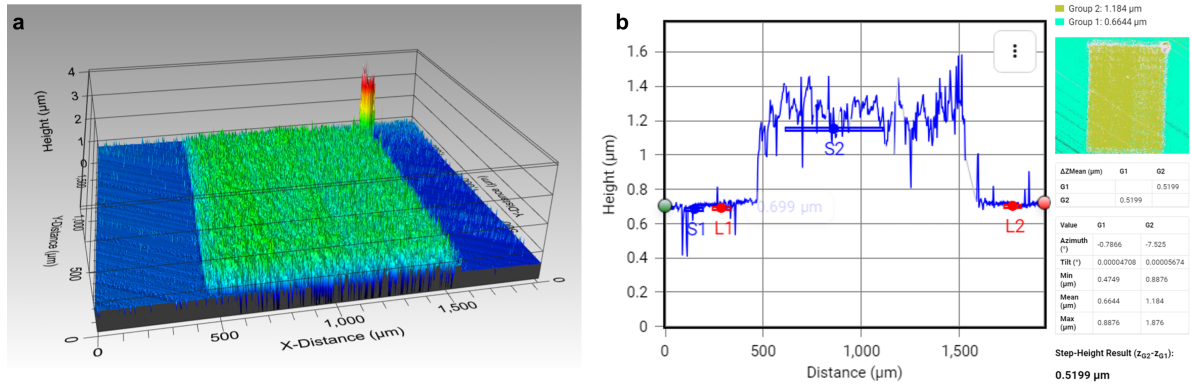

**Figure S2. Z-plane printing resolution.** (a) A typical white light profilometer scan across the printed 1-layer PUD film showing the deposition resolution as small as 500 nm (nozzle size: 25 gauge). (b) Surface topology of the printed film.

**Supplementary Note 3. Design of each gradient pattern**

**Figure S3** illustrates the dimensions of various Universal Gradient Interface (UGI) designs. The non-gradient design, used as a reference, features an abrupt modulus transition along the X-direction. The 1D in-plane gradient design exhibits a gradual modulus change in the X-direction. Conversely, the 1D out-of-plane gradient design transitions modulus changes along the Z-direction. The 2D in-plane gradient design modifies modulus progressively in both the X and Y directions. Lastly, the 3D gradient design systematically alters the modulus across all three dimensions: X, Y, and Z.

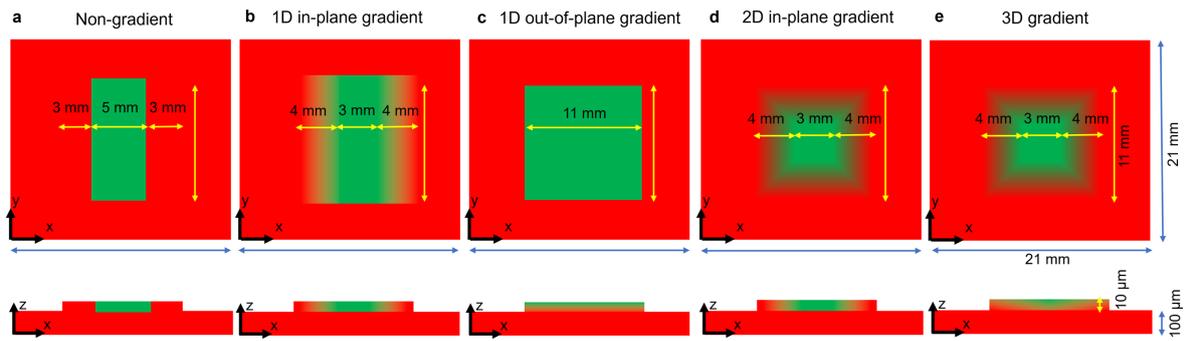

**Figure S3.** Design of each non-gradient and gradient pattern.

**Supplementary Note 4. Non-gradient and 1D out-of-plane UGI properties**
**Figures S4a** and **S4d** show the printed non-gradient and 1D out-of-plane interface pattern which are used for comparison purposes. **Figures S4b** and **S4e** show the site-specific component ratio of the non-gradient and 1D out-of-plane gradient design. The top surface principal Lagrange strain map obtained from DIC for non-gradient and 1D out-of-plane gradient samples under 100% global external stretch (**Figures S4c** and **S4f**) illustrates the non-gradient and 1D out-of-plane gradient samples already break at the soft-stiff interfaces.

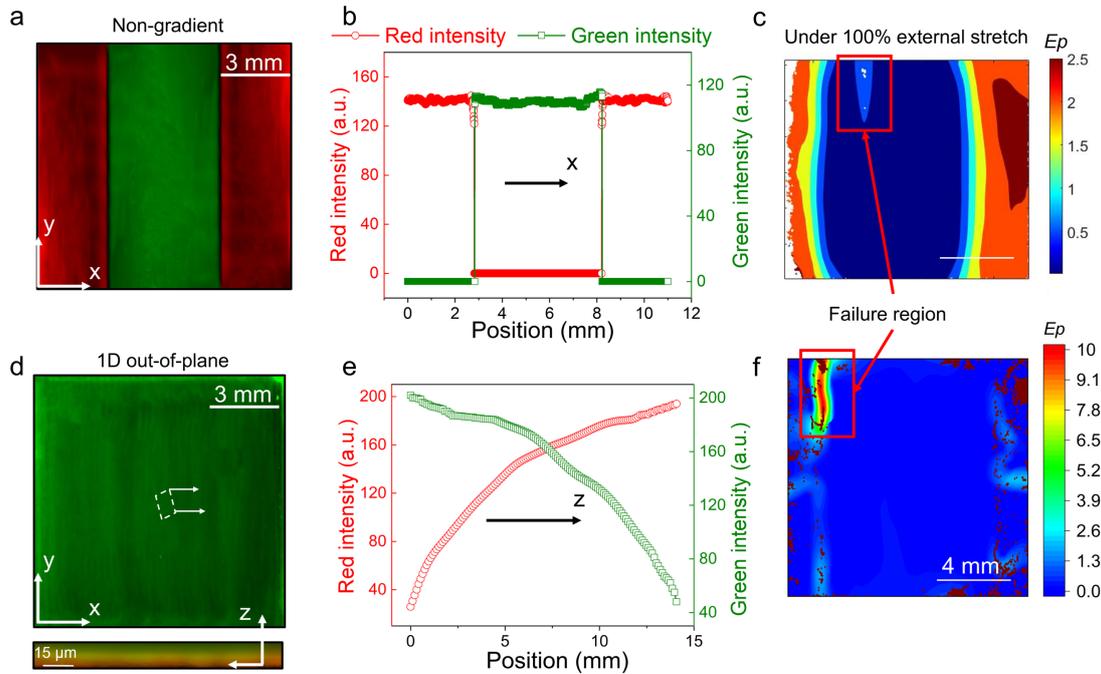

**Figure S4. Non-gradient and 1D out-of-plane UGI properties.** (a) Printed pattern, (b) site-specific component ratio, and (c) DIC under 100% stretch ratio for non-gradient design. (d) Printed pattern, (e) site-specific component ratio, and (f) DIC under 100% stretch ratio for 1D out-of-plane gradient design.

## Supplementary Note 5. DIC figures for each design at crack

**Figure S5** shows the top surface maximum principal logarithmic strain map obtained from DIC for each design when a crack occurs.

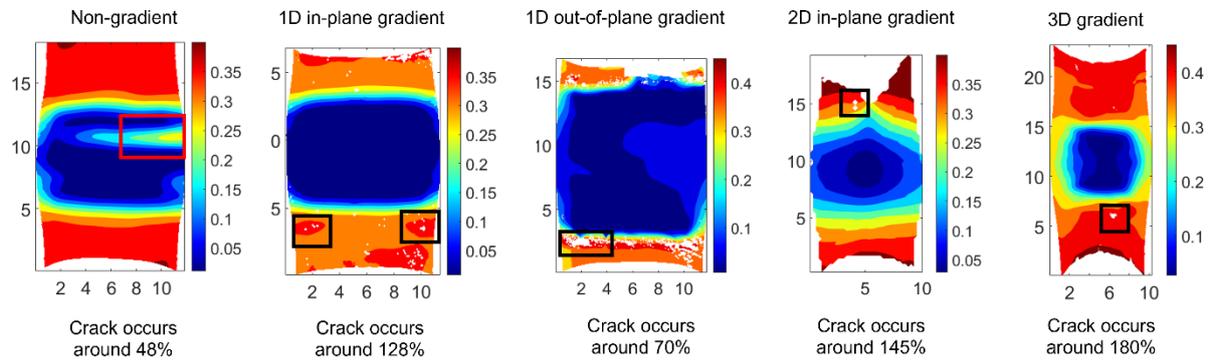

**Figure S5. DIC figures for each design at crack.**

**Supplementary Note 6. Stretching process for each design**

**Figure S6** shows the optical images of the stretching process of 0-100% stretch ratio and the images of cracks occurring in each UGI.

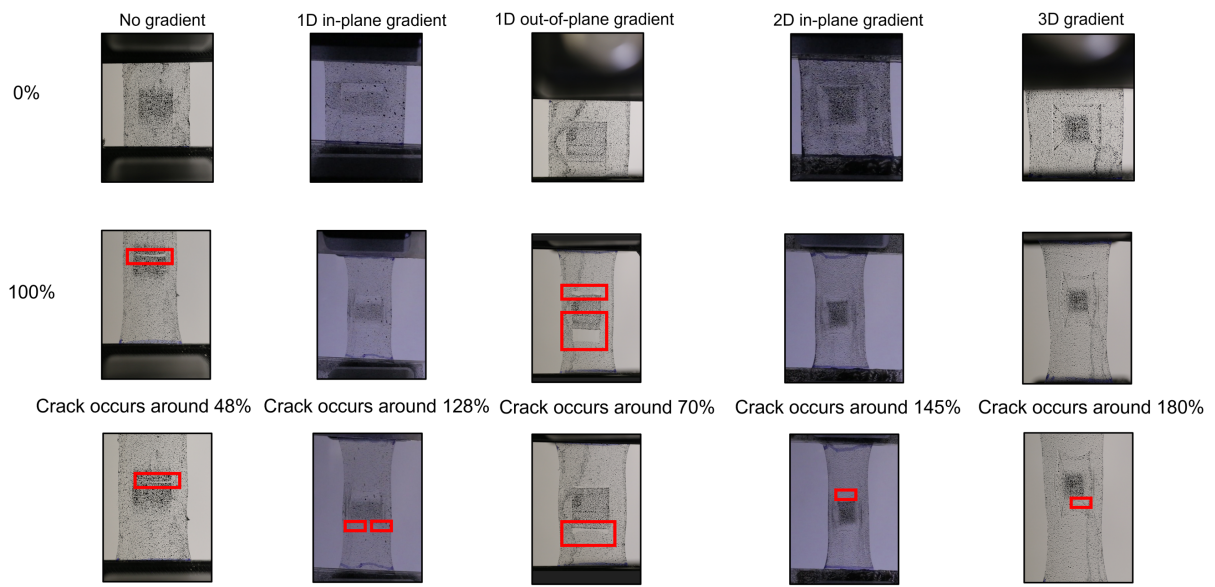

**Figure S6. Optical image of stretching process for each design.**

**Supplementary Note 7. Tensile test for non-gradient and 3D UGI design**

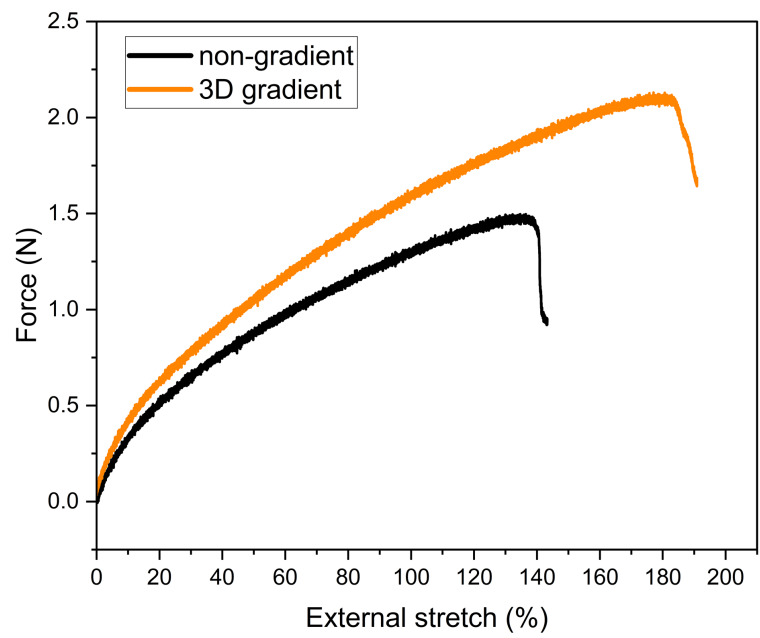

**Figure S7. Tensile test for non-gradient and 3D UGI design.**

**Supplementary Note 8. Simulation for each design for uniaxial stretch ratio of 100%**

**Figure S8** shows the simulation results of maximum principal strain and principal maximum stress under a uniaxial stretch ratio of 100% for each UGI. **Figures S8a** and **S8b** show the schematic and dimensions of the simulation process. By symmetry, only a quarter of the specimen is modeled with XSYMM and YSYMM boundary conditions at the left and bottom separately. **Figures. S8c** to **S8g** illustrate the maximum principal logarithmic strain distribution of each design. **Figures. S8h** to **S8l** show the principal maximum stress distribution of each design.

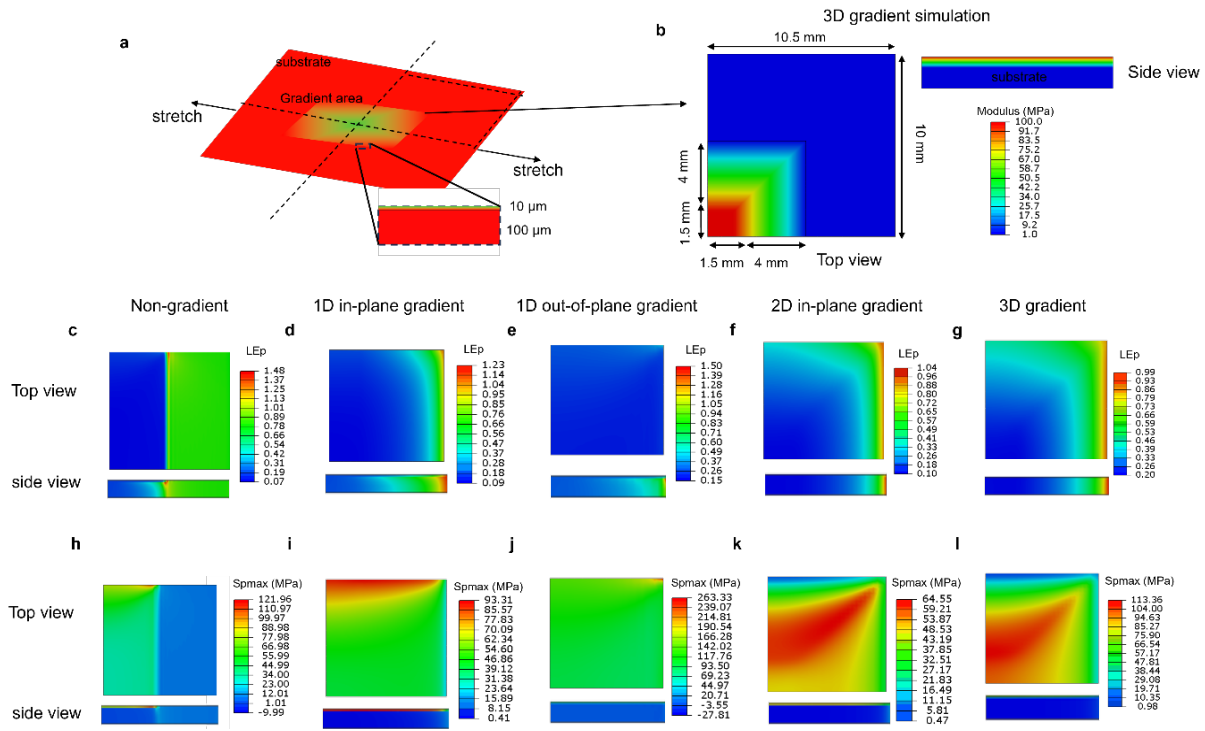

**Figure S8. Simulation for each design for uniaxial stretch ratio of 100%.** (a) Schematic of the simulation setup. (b) Simulation dimensions with 2D in-plane gradient design as an example. (c)-(g), Principal logarithmic strain distribution of each design. (h)-(l), Maximum principal stress distribution of each design.

**Supplementary Note 9. Simulation for each design for bi-axial stretch ratio of 100%**

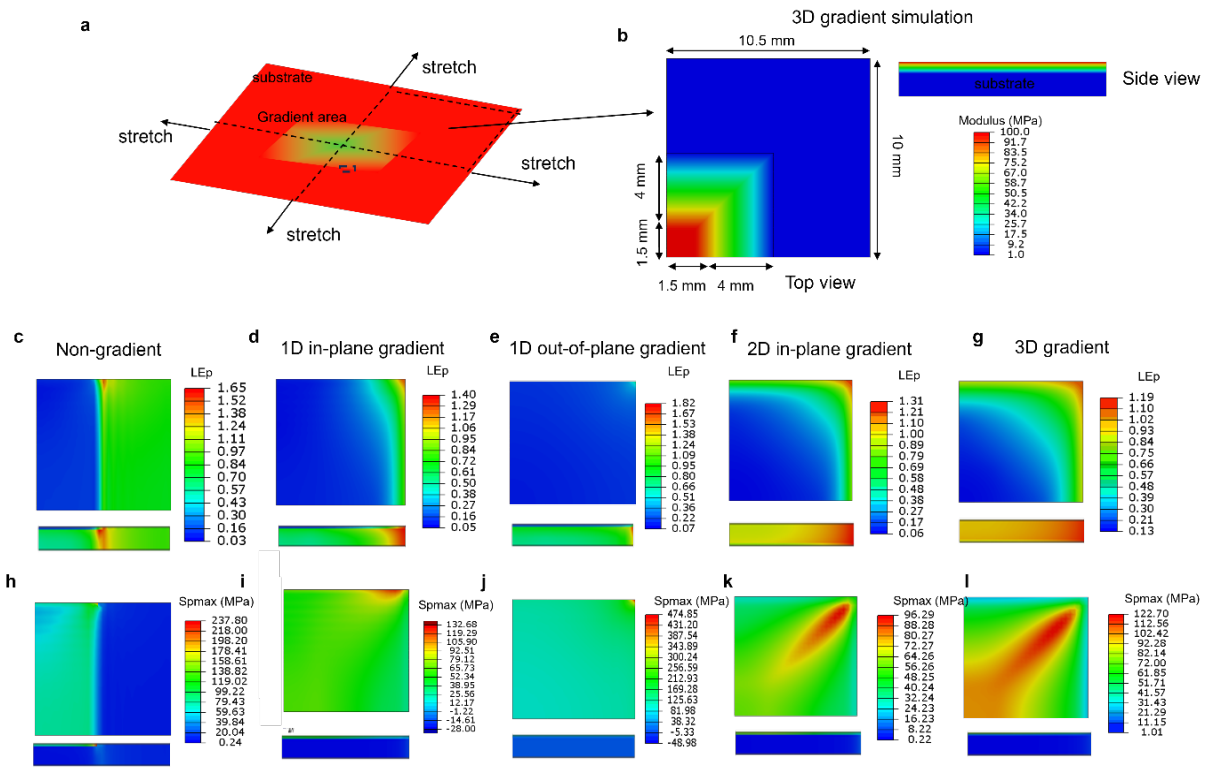

**Figure S9. Simulation for each design for bi-axial stretch ratio of 100%.** (a) Schematic of the simulation setup. (b) Simulation dimensions with 3D in-plane gradient design as an example. (c)-(g), Principal logarithmic strain distribution of each design. (h)-(l), Maximum principal stress distribution of each design.

**Supplementary Note 10. Flash sintering of printed Au sensor**

Flash sintering is one type of photonic sintering process where white flashlight energy is utilized to create local heating, resulting in the evaporation of solvents, integration of the nanoparticles being sintered, increasing the electrical and thermal conductivity, reducing porosity, etc. In this study, we used a Sinteron 2100 (Xenon Corporation, USA) integrated with a 107 mm Xenon spiral flash lamp. The xenon flash lamp generates white light by converting the electrical energy into light energy.[1] The short-pulsed white light from the xenon lamp covers a small part of UV light (380 nm), infrared light (950 nm), and the entire length of visible light.[2] Due to the surface plasmon resonance effect of the metal nanoparticles in the range of the visible light spectrum, when flash sinters metal nanoparticles, they absorb the light energy, resulting in local heating and completing the sintering process.[1] As the whole process is completed in a few seconds, the flash sintering is faster than other sintering processes, such as thermal sintering, laser sintering, etc. Targeted and localized sintering is also possible by flash sintering.

Here, the printed gold nanoparticles were sintered in air at 2 KV flash lamp voltage, 800 μs flashlight duration, and 3 consecutive pulses with a pulse delay of 858 ms, as shown in **Figure S10a**. As the optimized sintering time is less than 2s and the PUD substrate only absorbs the UV light of the xenon flash lamp,[1] metal nanoparticles can be sintered on a PUD substrate without potentially damaging it. The comparison of the printed gold filaments before and after flash sintering are illustrated in **Figures S10b** and **S10c**.

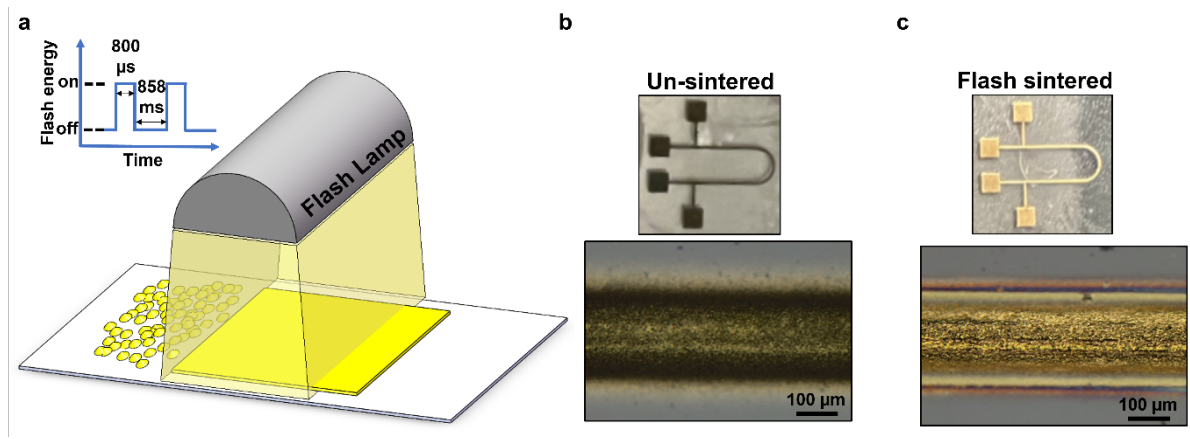

**Figure S10. Flash sintering of printed Au sensor.** (a) ultra-fast flash sintering of the printed gold and the sintering condition. Printed gold (b) before and (c) after flash sintering.

**Supplementary Note 11. Stability curve of the printed MXene-based sensor**

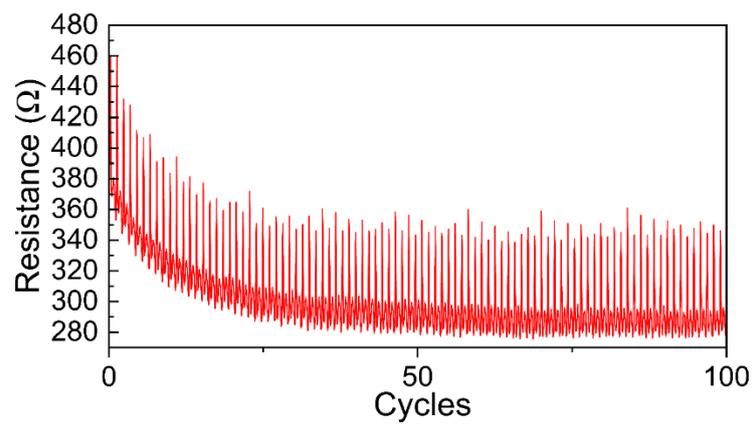

**Figure S11.** The stability curve of the printed MXene-based sensor on soft PUD substrate is under 20% strain for 100 cycles.

**Supplementary Note 12. Au-based temperature sensor**

Here the detailed design and fabrication process of the stretchable temperature sensor are illustrated in **Figure S12a**. **Figure S12b** shows the resistance change of the temperature sensor accurately following the changed temperature, which demonstrates the accuracy of the temperature sensor.

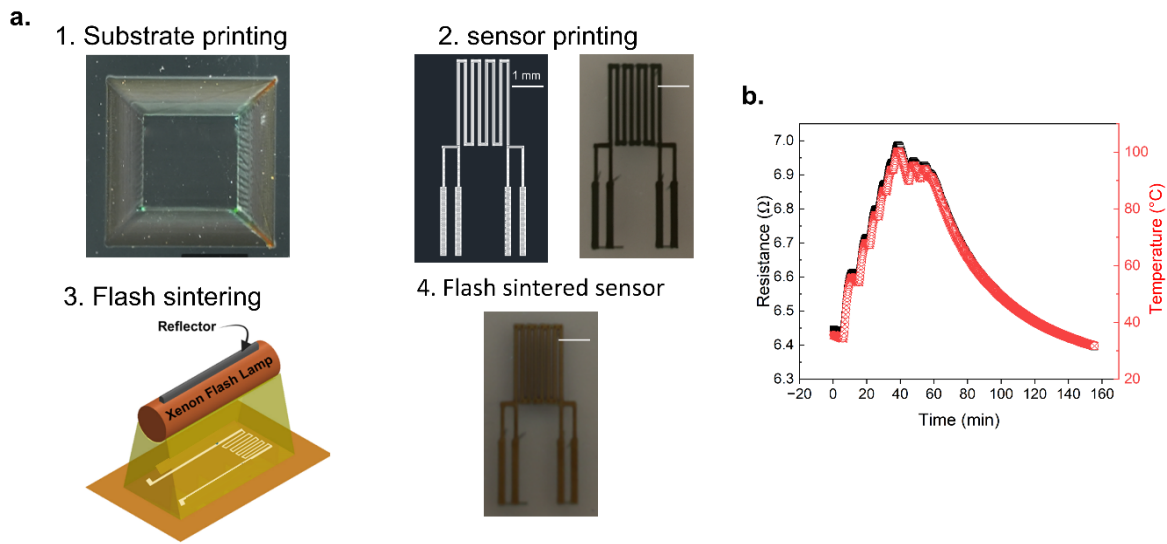

**Figure S12.** (a) The design and fabrication process of the temperature sensor. (b) The curve of resistance of the sensor under different temperatures.

**Supplementary Note 13. Substrate transparency analysis**

Here the transparency of PUD substrate under different wavelengths is illustrated in **Figure S13**. The high transparency confirms its capability for optical sensor applications.

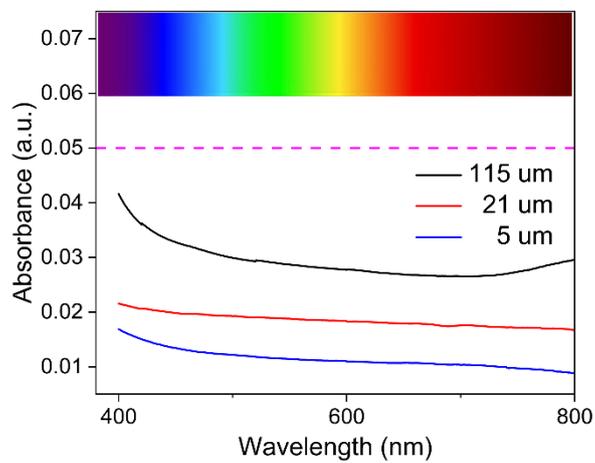

**Figure S13. Substrate transparency analysis**.

**Supplementary Note 14. Optical sensor fabrication process**

The MoS$_2$-based optical sensor or photodetector fabrication process is illustrated here. First, the MoS$_2$ was printed on gradient substrate. Then the MoS$_2$ layer was sintered in air at 2.4 KV flash lamp voltage, 800 μs flashlight duration, and 2 consecutive pulses with a pulse delay of 1360 ms through flashlight sintering. At last, MXene were printed on MoS$_2$ as electrodes.

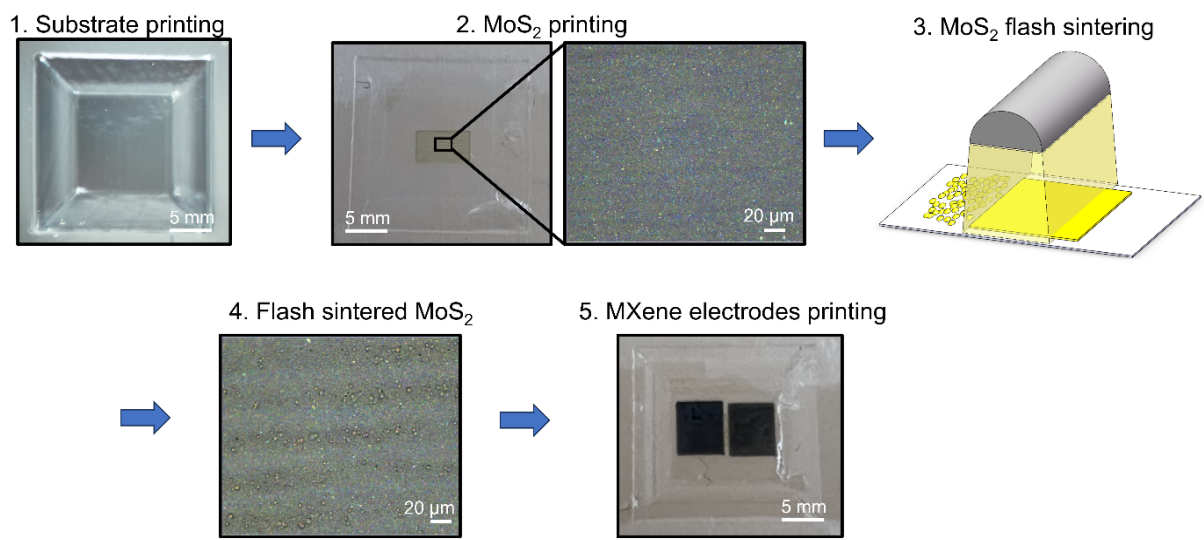

**Figure S14. Optical sensor fabrication process**.

**Supplementary Note 15. Responsivity of printed optical sensor**

For measuring the photodetector responsivity, we can assume that the intensity is uniform throughout the device channel since the laser spot size is substantially larger than the channel and covers the entire device. The total power was determined by multiplying the channel area by the intensity at the set laser current. The effective channel area is approximately 300 μm x 300 μm, and the input 405 nm laser has a diameter of 460 μm. Thus, we can calculate responsivity, $R = I_{pc}/P$, where $I_{pc}$ is the photocurrent and $P$ is the laser input power. The high responsivity observed is primarily due to the applied voltage across the photodetector, enhancing the electric field which in turn affects charge carrier dynamics. This applied voltage improves photocurrent generation by decreasing carrier recombination and boosting light-to-electricity conversion efficiency. In $MoS_2$ photoresistors, which operate based on the photoconductive principle, such voltage application not only increases conductivity with light exposure but also facilitates charge separation, further amplifying the photocurrent and thereby enhancing responsivity.

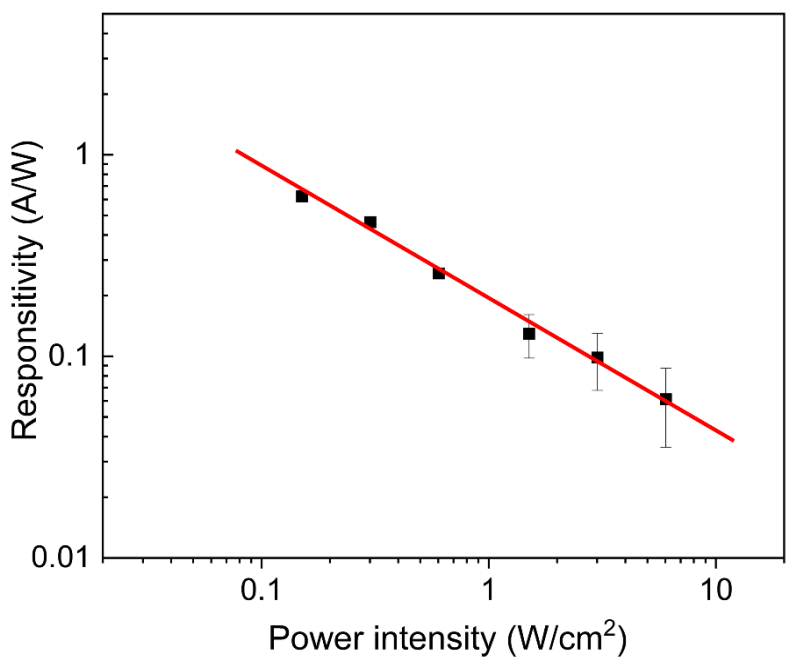

**Figure S15. Responsivity as a function of illumination power measured with a 405 nm laser under a bias voltage of 5V.**

**Supplementary Note 16. Response time and responsivity relationship of printed optical sensor**

**Figure S16a** shows the relationship between the printed MoS$_2$ thickness with response time. A thinner layer of MoS$_2$ in photodetectors generally leads to a faster response time primarily due to the reduced volume of material through which charge carriers need to move. While too thinner MoS$_2$ layers might not absorb light as effectively due to their reduced thickness, which can decrease the generation rate of electron-hole pairs. This reduced generation rate means fewer carriers are available to contribute to the photocurrent, potentially increasing the time it takes to reach a detectable signal level. **Figure S16b** shows the comparison of responsivity and response time in the devices in this work with previously reported all-printed photodetectors in the visible light range.

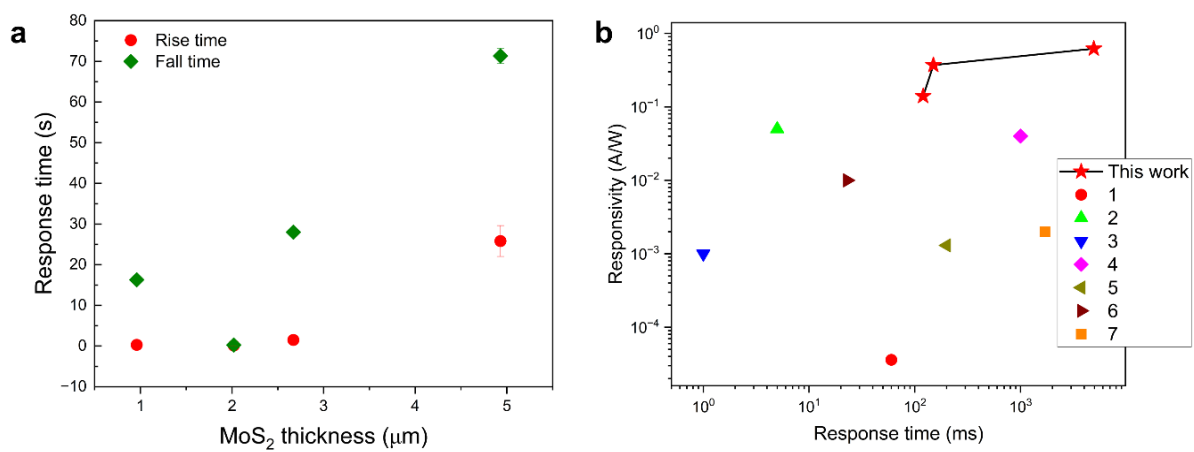

**Figure S16.** (a) Detailed response time and responsivity relationship. (b) Comparison of responsivity and response time in the devices in this work with previously reported all-printed photodetectors in the visible light range.

**Supplementary Note 17. Setup for optical sensor array application**

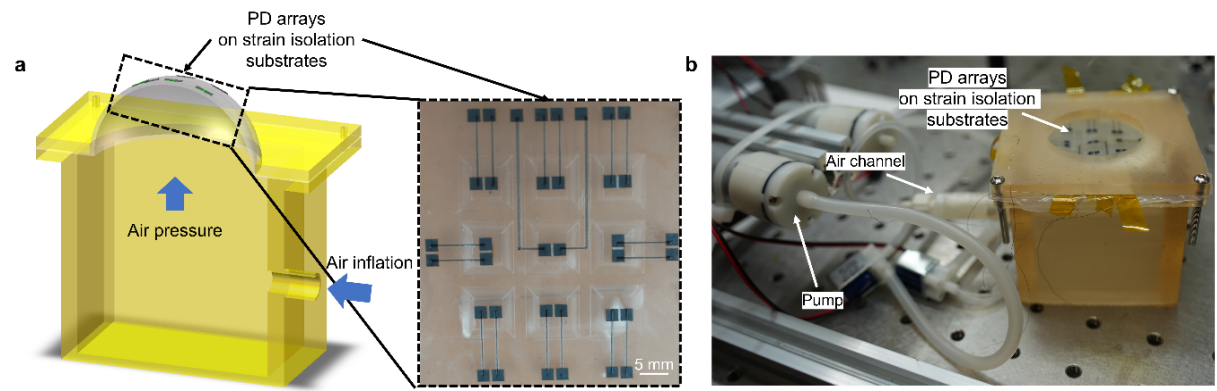

**Figure S17.** (a) Schematic illustration of the cross-sectional view of the pressure-tunable cylindrical cavity. (b) Optical image of the pressure-tunable cylindrical cavity connected with a mini pump.

**Supplementary Note 18. Simualtion of curved surface deformation**
**Figure S18a** illustrates the dimensions of the surface being simulated. Owing to symmetry, only one-quarter of the specimen is modeled, as depicted in **Figure S18b**. **Figure S18c** presents the simulated results of the strain distribution.

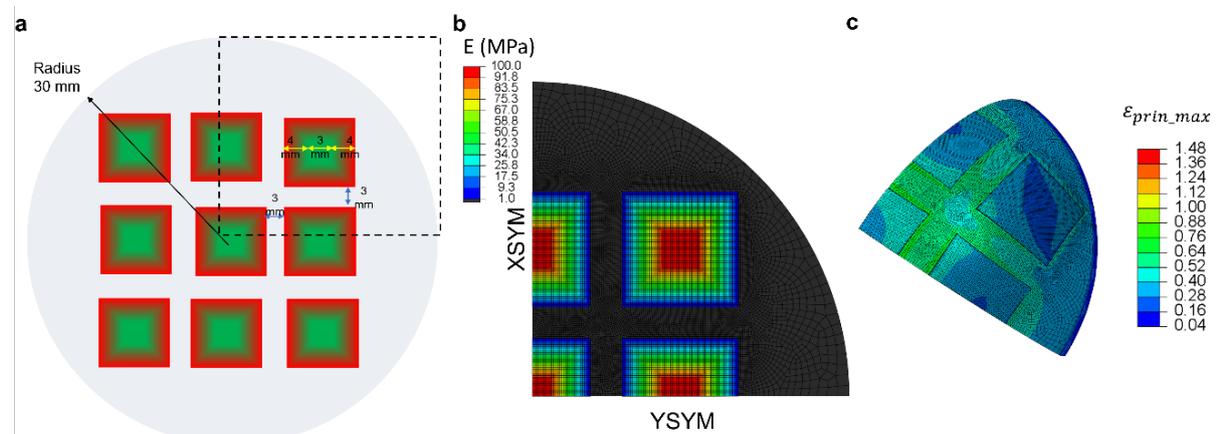

**Figure S18.** (a) Schematic illustration of dimension for simulation. (b) Modulus distribution of the sample. (c) Strain distribution during stretching.

**Supplementary Note 19. 3D photodetector array location coding and photocurrent**

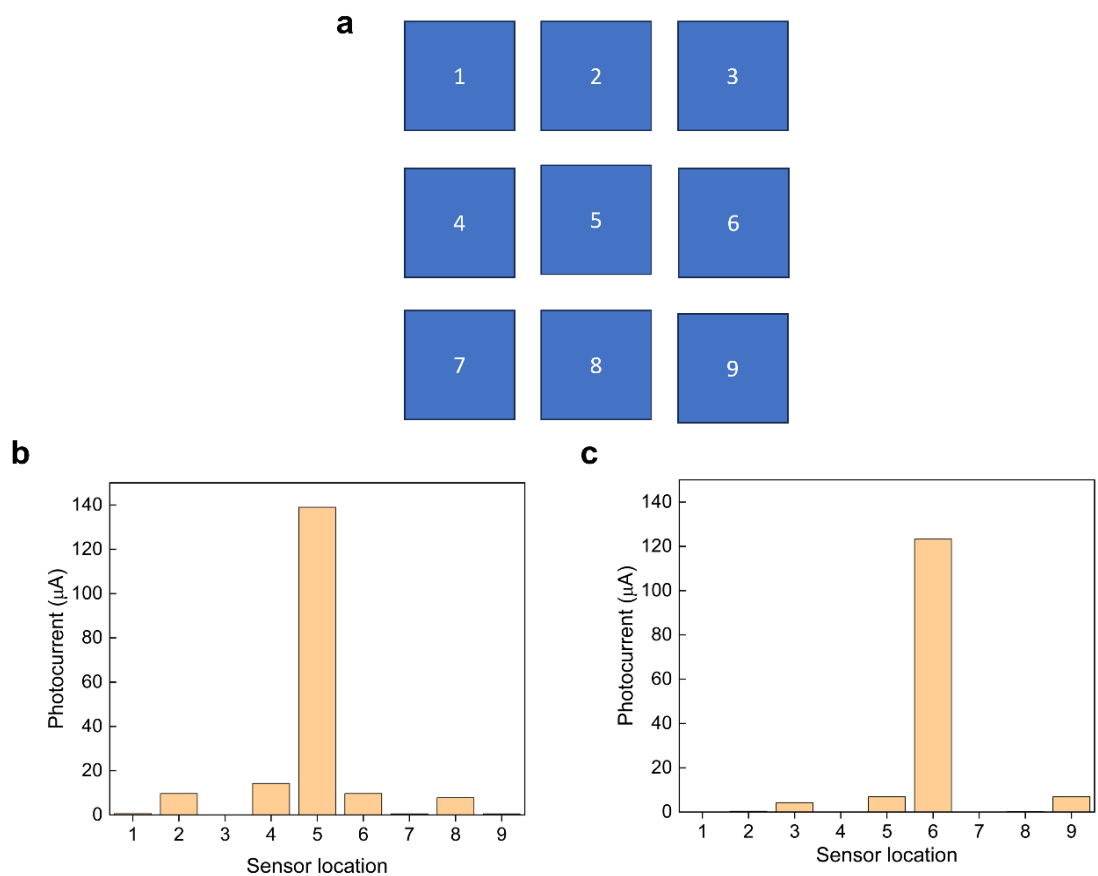

**Figure S19.** (a) 3D photodetector array location coding. Photocurrent of printed 3D photodetector array (b) under normally-incident illumination and (c) under obliquely incident light.

**Supplementary Note 20. Layout of whole wearable multi-model device**

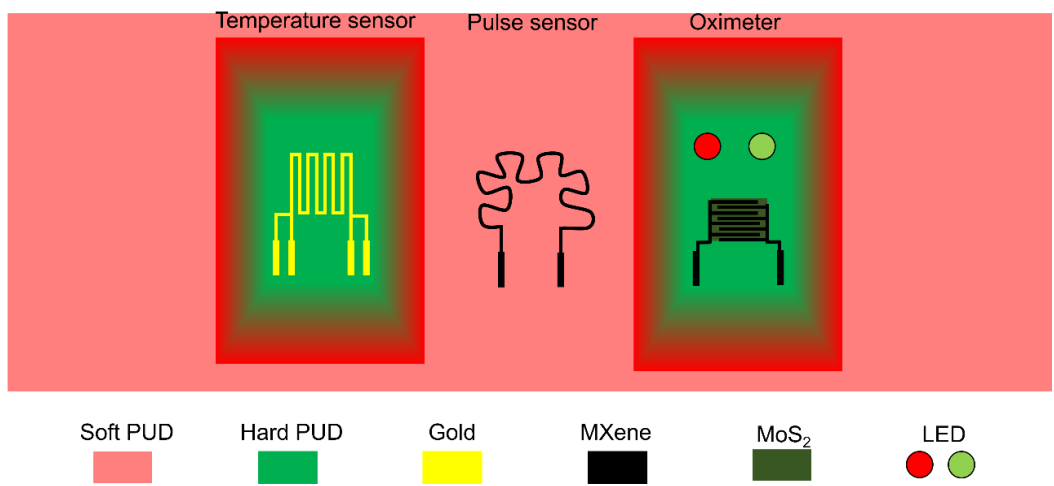

**Figure S20. Layout of whole wearable multi-model device.**

**Supplementary Note 21. Oxymeter calculation**

Blood oxygen saturation (SO$_2$) is quantified according to **Equation S1**. Here, C$_{HbO2}$ and C$_{Hb}$ are the concentrations of oxy-hemoglobin and deoxy-hemoglobin, respectively.

$$SO_2 = \frac{C_{HBO2}}{C_{HBO2}+C_{HB}} \quad (S1)$$

In this context, we use the characteristic value "Ros", the ratio of signal from red (A$_{rd}$) and green (A$_{gr}$) light according to Beer–Lambert's law (shown in **Equation S2**) to represent the blood oxygen saturation (SO$_2$):

$$R_{os} = \frac{A_{rd}}{A_{gr}} \approx \frac{AC_{rd}/DC_{rd}}{AC_{gr}/DC_{gr}} \quad (S2)$$

AC$_{rd}$ and DC$_{rd}$, along with AC$_{gr}$ and DC$_{gr}$, represent signals derived from red (650 nm) and green (530 nm) light PPG waveforms, respectively. As demonstrated in **Figures S21a**, these signals can be categorized into DC and AC components when converting light signals to electrical signals. The DC component is relatively constant across non-arterial tissues like muscles, bones, veins, and connective tissues, provided there is minimal movement at the measurement site. In contrast, the AC component varies due to dynamic changes in arterial light absorption, capturing fluctuations in arterial blood volume linked to cardiac cycles, specifically the phases of diastole and systole.

Direct comparison of red and green light signals may be unreliable due to variations in their light sources. The disparity in the DC component stems from the differing absorption rates of red and green light by skin tissue. To ensure equitable analysis of PPG signals from both light sources, the DC components are typically normalized through proportional calculations, as depicted in **Figures. S21b** and **S21c**. This normalization facilitates straightforward calculation of the R$_{os}$ value:

$$R_{os} \approx \frac{AC_{rd}}{AC_{gr}} \quad (S3)$$

For illustration, we make the measurement during a breath hold process, where SO$_2$ levels affect the oximeter readings. At an SO$_2$ of 99%, the AC$_{rd}$ is 0.135 µA, and AC$_{gr}$ is 0.150 µA, yielding a R$_{os}$ value of 0.9. When SO$_2$ drops to 96%, AC$_{rd}$ increases to 0.178 µA while AC$_{gr}$ decreases to 0.134 µA, resulting in a R$_{os}$ of 1.328. This demonstrates a significant rise in R$_{os}$ correlating with decreases in SO$_2$ during breath hold process. Consequently, our oximeter effectively measures and interprets real-time, non-invasive oxygen saturation, facilitating ongoing, discreet surveillance of physiological changes for early anomaly detection and disease prevention.

According to the Beer-Lambert Law and prior research,[3,4] SO$_2$ is determined by **Equation S4**, which directly correlates SO$_2$ with the R$_{os}$ value. The terms ε$_{rd,Hb}$ and ε$_{gr,Hb}$ represent the molar absorptivities of deoxy-hemoglobin at the red ($\lambda$ = 626 nm) and green ($\lambda$ = 530 nm) wavelengths, respectively. In a similar vein, ε$_{rd;HbO2}$ and ε$_{gr;HbO2}$ denote the molar absorptivities of oxy-hemoglobin at these same red and green wavelengths, respectively.

$$SO_2(R_{os}) = \frac{\varepsilon_{rd,Hb} - \varepsilon_{gr,Hb} R_{os}}{(\varepsilon_{rd,Hb} - \varepsilon_{rd,HbO_2}) + (\varepsilon_{gr,Hb} - \varepsilon_{gr,HbO_2}) R_{os}} \quad (S4)$$

To address the limitations of the Beer-Lambert Law, empirical corrections are necessary. By aligning the collected $R_{os}$ values with the $SO_2$ values from a commercial sensor, a calibration curve is presented in **Figure S21d**.

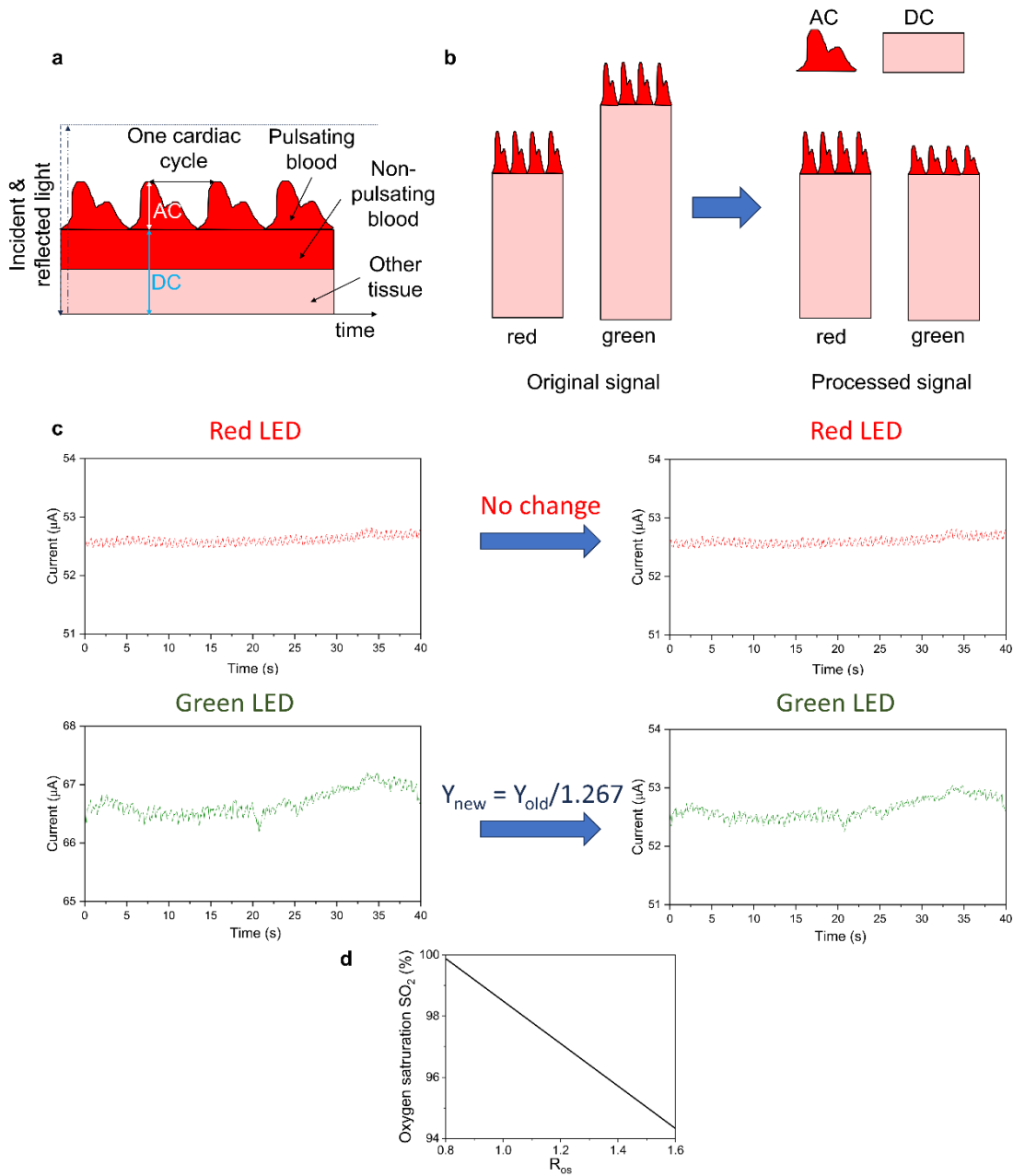

**Figure S21.** (a) Composition of AC and DC components in the PPG waveform. (b) Schematic representation of both original and processed signals for red (650 nm) and green (530 nm) light. (c) Processing diagram used to mitigate the influence of DC components. (d) Graph of matching $R_{os}$ values with $SO_2$.

## Supplementary Note 22. On skin performance

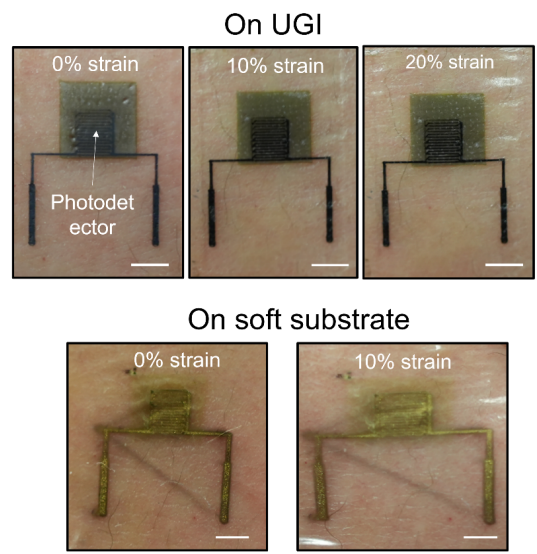

**Figure S22.** Oximeter printed on UGI and soft substrate before and after on-skin stretching. Scale bar, 2.5 mm.

**Table S1.** Typical stretchable materials and their performance.

| Ref | Plot number in main content | Resistance increase | Methods |
|---|---|---|---|
| [5] | 9 | The resistance of the elastic conductors remained the same under high bending strain and increased slightly (≈ 14%) at a tensile strain of 100% | Apply pre-strain |
| [6] | 27 | Larger pre-strain (20, 30, 40%) resulted in stable conductance | Apply pre-strain |
| [7] | 31 | Stable of stretching at ε = 0.7 | Intrinsically stretchable material |
| [8] | 32 | Resistance maintained at 60% (longitudinal), 140% (transverse) strain | Intrinsically stretchable material |
| [9] | 33 | The resistance keeps increasing slightly | Intrinsically stretchable material |
| [10] | 34 | Conductivity maintained up to 50% strain | Wrinkle and Serpentine |
| [11] | 35 | ΔR/R0 = 0.1 at 80% strain | Wrinkle |
| [12] | 36 | Conductivity maintained up to 50% strain | Serpentine |
| [13] | 37 | Resistance slightly increased up to 200% strain | Serpentine |
| [14] | 38 | Conductivity maintained up to 100% strain | Serpentine |
| [15] | 39 | Resistance slightly increased up to 200% strain | Rigid island |
| [16] | 40 | The resistance keeps increasing slightly | Gradient substrate |
| [17] | 41 | The resistance keeps increasing slightly | Rigid island |

**Table S2.** Literature comparison for Fig. S16b.

| Ref | Plot label number | Printing method | Response time (ms) | Responsivity (A/W) |
|---|---|---|---|---|
|  | This work | AJP | 120 | 0.14 |
|  | This work | AJP | 150 | 0.37 |
|  | This work | AJP | 4930 | 0.62 |
| [18] | 1 | Inkjet | 60 | 3.6 x $10^{-5}$ |
| [19] | 2 | Inkjet | 5 | 0.05 |
| [20] | 3 | Inkjet | 1 | 0.001 |
| [21] | 4 | Electrohydrodynamic | 1000 | 0.004 |
| [22] | 5 | Laser printing | 200 | 0.0013 |
| [23] | 6 | Inkjet | 23 | 0.01 |
| [24] | 7 | Inkjet | 1700 | 0.002 |

**Table S3.** Typical motion artifact-free devices' performance.

| Ref | Sensor type | Percent signal variance/drift | Applied strain |
|---|---|---|---|
| This work | Oximeter | <1.7% | ≈ 20% |
| [25] | Hydration sensor | ≈ 14.7% | ≈ 10% |
| [26] | Epidermal antenna | ≈ 80% | ≈ 20% |
| [27] | EMG electrode array | ≈ 60% | Unknown |
| [28] | Solid-state analyte sensor | ≈ 21% | Unknown |
| [29] | Pulse sensor | ≈ 50% | Unknown |

**Table S4.** Typical ink formulation of nanoparticle inks.

| Materials | Solvent | Co-solvent | Mass percentage | Surfactants/Additives |
|---|---|---|---|---|
| Soft PUD | Water | DMSO | 30% | EG |
| Hard PUD | Water | | 27.3% | EG |
| PEDOT: PSS | Water | | 0.8% | |
| Gold | Xylene | | 2% | |
| AgNWs | Water | | 1% | |
| MXene | Water | | 0.8% | |
| $MoS_2$ | IPA | | 0.6% | Terpinol |

**Table S5.** Aerosol jet printing parameters for each gradient substrate design.

| Parameters | Non-gradient | Linear gradient | Z-gradient | XY-gradient | XYZ-gradient |
|---|---|---|---|---|---|
| Nozzle nominal I.D. (μm) | 410 | 410 | 410 | 410 | 410 |
| Sheath gas flow rate (sccm) | 60 | 60 | 60 | 60 | 60 |
| Soft PUD ink flow rate (sccm) | 30 | 10-30 | 10-30 | 10-30 | 10-30 |
| Hard PUD ink flow rate (sccm) | 30 | 17-30 | 17-30 | 17-30 | 17-30 |
| Platen temperature (°C) | 50 | 50 | 50 | 50 | 50 |
| Print speed (mm/s) | 2 | 2 | 2 | 2 | 2 |
| Ultrasonic atomizing voltage (V) (soft) | 43 | 43 | 43 | 43 | 43 |
| Ultrasonic atomizing voltage (V)(hard) | 43 | 43 | 43 | 43 | 43 |

**Table S6.** Aerosol jet printing parameters for each sensor.

| Parameters | Strain sensor | Temperature sensor | Oximeter (channel) | Oximeter (electrode) |
|---|---|---|---|---|
| Nozzle nominal I.D. (μm) | 250 | 250 | 410 | 250 |
| Sheath gas flow rate (sccm) | 80 | 80 | 60 | 80 |
| Ink flow rate (sccm) | 10 | 10 | 15 | 12 |
| Platen temperature (°C) | 50 | 50 | 50 | 50 |
| Print speed (mm/s) | 2 | 2 | 2 | 2 |
| Ultrasonic atomizing voltage (V) | 43 | 43 | 43 | 43 |

**References**

[1] G. L. Goh, H. Zhang, T. H. Chong, W. Y. Yeong, *Adv Electron Mater* **2021**, *7*, 1.

[2] Y. R. Jang, S. J. Joo, J. H. Chu, H. J. Uhm, J. W. Park, C. H. Ryu, M. H. Yu, H. S. Kim, *A Review on Intense Pulsed Light Sintering Technologies for Conductive Electrodes in Printed Electronics*, Korean Society For Precision Engineering, **2021**.

[3] C. M. Lochner, Y. Khan, A. Pierre, A. C. Arias, *Nat Commun* **2014**, *5*, 1.

[4] H. Lee, W. Lee, H. Lee, S. Kim, M. V. Alban, J. Song, T. Kim, S. Lee, S. Yoo, *ACS Photonics* **2021**, *8*, 3564.

[5] K. H. Kim, M. Vural, M. F. Islam, *Advanced Materials* **2011**, *23*, 2865.

[6] A. Miyamoto, S. Lee, N. F. Cooray, S. Lee, M. Mori, N. Matsuhisa, H. Jin, L. Yoda, T. Yokota, A. Itoh, M. Sekino, H. Kawasaki, T. Ebihara, M. Amagai, T. Someya, *Nat Nanotechnol* **2017**, *12*, 907.

[7] M. Shin, J. H. Song, G. H. Lim, B. Lim, J. J. Park, U. Jeong, *Advanced Materials* **2014**, *26*, 3706.

[8] G. D. Moon, G. H. Lim, J. H. Song, M. Shin, T. Yu, B. Lim, U. Jeong, *Advanced Materials* **2013**, *25*, 2707.

[9] P. Lee, J. Lee, H. Lee, J. Yeo, S. Hong, K. H. Nam, D. Lee, S. S. Lee, S. H. Ko, *Advanced Materials* **2012**, *24*, 3326.

[10] D. C. Hyun, M. Park, C. Park, B. Kim, Y. Xia, J. H. Hur, J. M. Kim, J. J. Park, U. Jeong, *Advanced Materials* **2011**, *23*, 2946.

[11] V. R. Feig, H. Tran, M. Lee, Z. Bao, *Nat Commun* **2018**, *9*, 1.

[12] J. Park, S. Choi, A. H. Janardhan, S. Y. Lee, S. Raut, J. Soares, K. Shin, S. Yang, C. Lee, K. W. Kang, H. R. Cho, S. J. Kim, P. Seo, W. Hyun, S. Jung, H. J. Lee, N. Lee, S. H. Choi, M. Sacks, N. Lu, M. E. Josephson, T. Hyeon, D. H. Kim, H. J. Hwang, *Sci Transl Med* **2016**, *8*, 1.

[13] R. Ma, B. Kang, S. Cho, M. Choi, S. Baik, *ACS Nano* **2015**, *9*, 10876.

[14] S. Choi, J. Park, W. Hyun, J. Kim, J. Kim, Y. B. Lee, C. Song, H. J. Hwang, J. H. Kim, T. Hyeon, D. H. Kim, *ACS Nano* **2015**, *9*, 6626.

[15] R. Libanori, R. M. Erb, A. Reiser, H. Le Ferrand, M. J. Süess, R. Spolenak, A. R. Studart, *Nat Commun* **2012**, *3*, 1.

[16] T. Yang, Y. Zhong, D. Tao, X. Li, X. Zang, S. Lin, X. Jiang, Z. Li, H. Zhu, *2d Mater* **2017**, *4*, DOI 10.1088/2053-1583/aa78cc.

[17] Y. Zhao, B. Wang, J. Tan, H. Yin, R. strain-insensitive bioelectronics featuring brittle materials Huang, J. Zhu, S. Lin, Y. Zhou, D. Jelinek, Z. Sun, K. Youssef, L. Voisin, A. Horrillo, K. Zhang, B. M. Wu, H. A. Coller, D. C. Lu, Q. Pei, S. Emaminejad, *Science* **2022**, *378*, 1222.

[18] J. Li, M. M. Naiini, S. Vaziri, M. C. Lemme, M. Östling, *Adv Funct Mater* **2014**, *24*, 6524.

[19] J. W. T. Seo, J. Zhu, V. K. Sangwan, E. B. Secor, S. G. Wallace, M. C. Hersam, *ACS Appl Mater Interfaces* **2019**, *11*, 5675.

[20] D. McManus, S. Vranic, F. Withers, V. Sanchez-Romaguera, M. Macucci, H. Yang, R. Sorrentino, K. Parvez, S. K. Son, G. Iannaccone, K. Kostarelos, G. Fiori, C. Casiraghi, *Nat Nanotechnol* **2017**, *12*, 343.

[21]  F. I. Alzakia, W. Jonhson, J. Ding, S. C. Tan, *ACS Appl Mater Interfaces* **2020**, *12*, 28840.
[22]  A. Mazaheri, M. Lee, H. S. J. Van Der Zant, R. Frisenda, A. Castellanos-Gomez, *Nanoscale* **2020**, *12*, 19068.
[23]  L. Kong, G. Li, Q. Su, X. Zhang, Z. Liu, G. Liao, B. Sun, T. Shi, *Adv Eng Mater* **2023**, *25*, 1.
[24]  T. Y. Kim, J. Ha, K. Cho, J. Pak, J. Seo, J. Park, J. K. Kim, S. Chung, Y. Hong, T. Lee, *ACS Nano* **2017**, *11*, 10273.
[25]  F. Ershad, A. Thukral, J. Yue, P. Comeaux, Y. Lu, H. Shim, K. Sim, N. I. Kim, Z. Rao, R. Guevara, L. Contreras, F. Pan, Y. Zhang, Y. S. Guan, P. Yang, X. Wang, P. Wang, X. Wu, C. Yu, *Nat Commun* **2020**, *11*, 1.
[26]  Z. Wang, X. Xiao, W. Wu, X. Zhang, Y. Pang, *Biosens Bioelectron* **2024**, *253*, 116150.
[27]  Y. Jiang, S. Ji, J. Sun, J. Huang, Y. Li, G. Zou, T. Salim, C. Wang, W. Li, H. Jin, J. Xu, S. Wang, T. Lei, X. Yan, W. Y. X. Peh, S. C. Yen, Z. Liu, M. Yu, H. Zhao, Z. Lu, G. Li, H. Gao, Z. Liu, Z. Bao, X. Chen, *Nature* **2023**, *614*, 456.
[28]  R. T. Arwani, S. C. L. Tan, A. Sundarapandi, W. P. Goh, Y. Liu, F. Y. Leong, W. Yang, X. T. Zheng, Y. Yu, C. Jiang, Y. C. Ang, L. Kong, S. L. Teo, P. Chen, X. Su, H. Li, Z. Liu, X. Chen, L. Yang, Y. Liu, *Nat Mater* **2024**, *23*, DOI 10.1038/s41563-024-01918-9.
[29]  C. Jeong, G. R. Koirala, Y. H. Jung, Y. S. Ye, J. H. Hyun, T. H. Kim, B. Park, J. Ok, Y. Jung, T. il Kim, *Adv Funct Mater* **2022**, *32*, 1.

# References

[1] G. L. Goh, H. Zhang, T. H. Chong, W. Y. Yeong, *Adv Electron Mater* **2021**, *7*, 1.
[2] Y. R. Jang, S. J. Joo, J. H. Chu, H. J. Uhm, J. W. Park, C. H. Ryu, M. H. Yu, H. S. Kim, *A Review on Intense Pulsed Light Sintering Technologies for Conductive Electrodes in Printed Electronics*, Korean Society For Precision Engineering, **2021**.
[3] C. M. Lochner, Y. Khan, A. Pierre, A. C. Arias, *Nat Commun* **2014**, *5*, 1.
[4] H. Lee, W. Lee, H. Lee, S. Kim, M. V. Alban, J. Song, T. Kim, S. Lee, S. Yoo, *ACS Photonics* **2021**, *8*, 3564.
[5] K. H. Kim, M. Vural, M. F. Islam, *Advanced Materials* **2011**, *23*, 2865.
[6] A. Miyamoto, S. Lee, N. F. Cooray, S. Lee, M. Mori, N. Matsuhisa, H. Jin, L. Yoda, T. Yokota, A. Itoh, M. Sekino, H. Kawasaki, T. Ebihara, M. Amagai, T. Someya, *Nat Nanotechnol* **2017**, *12*, 907.
[7] M. Shin, J. H. Song, G. H. Lim, B. Lim, J. J. Park, U. Jeong, *Advanced Materials* **2014**, *26*, 3706.
[8] G. D. Moon, G. H. Lim, J. H. Song, M. Shin, T. Yu, B. Lim, U. Jeong, *Advanced Materials* **2013**, *25*, 2707.
[9] P. Lee, J. Lee, H. Lee, J. Yeo, S. Hong, K. H. Nam, D. Lee, S. S. Lee, S. H. Ko, *Advanced Materials* **2012**, *24*, 3326.
[10] D. C. Hyun, M. Park, C. Park, B. Kim, Y. Xia, J. H. Hur, J. M. Kim, J. J. Park, U. Jeong, *Advanced Materials* **2011**, *23*, 2946.
[11] V. R. Feig, H. Tran, M. Lee, Z. Bao, *Nat Commun* **2018**, *9*, 1.
[12] J. Park, S. Choi, A. H. Janardhan, S. Y. Lee, S. Raut, J. Soares, K. Shin, S. Yang, C. Lee, K. W. Kang, H. R. Cho, S. J. Kim, P. Seo, W. Hyun, S. Jung, H. J. Lee, N. Lee, S. H. Choi, M. Sacks, N. Lu, M. E. Josephson, T. Hyeon, D. H. Kim, H. J. Hwang, *Sci Transl Med* **2016**, *8*, 1.
[13] R. Ma, B. Kang, S. Cho, M. Choi, S. Baik, *ACS Nano* **2015**, *9*, 10876.
[14] S. Choi, J. Park, W. Hyun, J. Kim, J. Kim, Y. B. Lee, C. Song, H. J. Hwang, J. H. Kim, T. Hyeon, D. H. Kim, *ACS Nano* **2015**, *9*, 6626.
[15] R. Libanori, R. M. Erb, A. Reiser, H. Le Ferrand, M. J. Süess, R. Spolenak, A. R. Studart, *Nat Commun* **2012**, *3*, 1.
[16] T. Yang, Y. Zhong, D. Tao, X. Li, X. Zang, S. Lin, X. Jiang, Z. Li, H. Zhu, *2d Mater* **2017**, *4*, DOI 10.1088/2053-1583/aa78cc.
[17] Y. Zhao, B. Wang, J. Tan, H. Yin, R. strain-insensitive bioelectronics featuring brittle materials Huang, J. Zhu, S. Lin, Y. Zhou, D. Jelinek, Z. Sun, K. Youssef, L. Voisin, A. Horrillo, K. Zhang, B. M. Wu, H. A. Coller, D. C. Lu, Q. Pei, S. Emaminejad, *Science* **2022**, *378*, 1222.
[18] J. Li, M. M. Naiini, S. Vaziri, M. C. Lemme, M. Östling, *Adv Funct Mater* **2014**, *24*, 6524.

[19] J. W. T. Seo, J. Zhu, V. K. Sangwan, E. B. Secor, S. G. Wallace, M. C. Hersam, *ACS Appl Mater Interfaces* **2019**, *11*, 5675.

[20] D. McManus, S. Vranic, F. Withers, V. Sanchez-Romaguera, M. Macucci, H. Yang, R. Sorrentino, K. Parvez, S. K. Son, G. Iannaccone, K. Kostarelos, G. Fiori, C. Casiraghi, *Nat Nanotechnol* **2017**, *12*, 343.

[21] F. I. Alzakia, W. Jonhson, J. Ding, S. C. Tan, *ACS Appl Mater Interfaces* **2020**, *12*, 28840.

[22] A. Mazaheri, M. Lee, H. S. J. Van Der Zant, R. Frisenda, A. Castellanos-Gomez, *Nanoscale* **2020**, *12*, 19068.

[23] L. Kong, G. Li, Q. Su, X. Zhang, Z. Liu, G. Liao, B. Sun, T. Shi, *Adv Eng Mater* **2023**, *25*, 1.

[24] T. Y. Kim, J. Ha, K. Cho, J. Pak, J. Seo, J. Park, J. K. Kim, S. Chung, Y. Hong, T. Lee, *ACS Nano* **2017**, *11*, 10273.

[25] F. Ershad, A. Thukral, J. Yue, P. Comeaux, Y. Lu, H. Shim, K. Sim, N. I. Kim, Z. Rao, R. Guevara, L. Contreras, F. Pan, Y. Zhang, Y. S. Guan, P. Yang, X. Wang, P. Wang, X. Wu, C. Yu, *Nat Commun* **2020**, *11*, 1.

[26] Z. Wang, X. Xiao, W. Wu, X. Zhang, Y. Pang, *Biosens Bioelectron* **2024**, *253*, 116150.

[27] Y. Jiang, S. Ji, J. Sun, J. Huang, Y. Li, G. Zou, T. Salim, C. Wang, W. Li, H. Jin, J. Xu, S. Wang, T. Lei, X. Yan, W. Y. X. Peh, S. C. Yen, Z. Liu, M. Yu, H. Zhao, Z. Lu, G. Li, H. Gao, Z. Liu, Z. Bao, X. Chen, *Nature* **2023**, *614*, 456.

[28] R. T. Arwani, S. C. L. Tan, A. Sundarapandi, W. P. Goh, Y. Liu, F. Y. Leong, W. Yang, X. T. Zheng, Y. Yu, C. Jiang, Y. C. Ang, L. Kong, S. L. Teo, P. Chen, X. Su, H. Li, Z. Liu, X. Chen, L. Yang, Y. Liu, *Nat Mater* **2024**, *23*, DOI 10.1038/s41563-024-01918-9.

[29] C. Jeong, G. R. Koirala, Y. H. Jung, Y. S. Ye, J. H. Hyun, T. H. Kim, B. Park, J. Ok, Y. Jung, T. il Kim, *Adv Funct Mater* **2022**, *32*, 1.